\documentclass[journal]{IEEEtran}
\usepackage{algorithmicx}
\usepackage{amsmath,amssymb,amsthm,cite}
\usepackage{graphicx}
\usepackage{mathtools}
\usepackage{color}
\usepackage[labelformat=simple]{subcaption}
\usepackage{pbox}
\usepackage{url}
\usepackage{bbm}
\usepackage{todonotes}
\usepackage[bookmarks=false,colorlinks=true,urlcolor=blue,citecolor=blue,linkcolor=blue]{hyperref}
\usepackage{cleveref}
\DeclareCaptionLabelSeparator{periodspace}{.\quad}
\newtheorem{lem}{Lemma}
\newtheorem{thm}{Theorem}
\newtheorem{defn}{Definition}
\newtheorem{coro}{Corollary}
\newtheorem{clm}{Claim}

\newtheorem{exple}{Example}

\let\oldexple\exple
\renewcommand{\exple}{\oldexple\normalfont}

\newcommand{\E}[1]{\mathbb{E}\left[{#1}\right]}

\newcommand{\HyperExp}{\textit{HyperExp}}

\newcommand{\Exp}{\textit{Exp}}
\newcommand{\Pareto}{\textit{Pareto}}

\crefname{equation}{}{}
\Crefname{equation}{}{}
\crefname{thm}{theorem}{theorems}
\Crefname{thm}{Theorem}{Theorems}
\crefname{clm}{claim}{claims}
\Crefname{clm}{Claim}{Claims}
\Crefname{coro}{Corollary}{Corollaries}
\Crefname{lem}{Lemma}{Lemmas}
\Crefname{sec}{Section}{Sections}
\crefname{app}{appendix}{appendices}
\Crefname{app}{Appendix}{Appendices}
\crefname{prop}{proposition}{propositions}
\Crefname{prop}{Proposition}{Propositions}
\Crefname{propty}{Property}{Properties}
\crefname{figure}{fig.}{figures}
\Crefname{figure}{Fig.}{Figures}
\crefname{defn}{definition}{definitions}
\Crefname{defn}{Definition}{Definitions}
\crefname{fact}{fact}{facts}
\Crefname{fact}{Fact}{Facts}
\crefname{appendix}{appendix}{appendices}
\Crefname{appendix}{Appendix}{Appendices}
\crefname{algo}{algorithm}{algorithms}
\Crefname{algo}{Algorithm}{Algorithms}
\crefname{algorithm}{algorithm}{algorithms}
\Crefname{algorithm}{Algorithm}{Algorithms}
\crefname{conj}{conjecture}{conjectures}
\Crefname{conj}{Conjecture}{Conjectures}
\crefname{obs}{observation}{observations}
\Crefname{obs}{Observation}{Observations}

\begin{document}

\newcommand{\NumServers}{K}
\newcommand{\Thpt}{R}
\newcommand{\Latency}{T}
\newcommand{\Cost}{C}
\newcommand{\CombTime}{S}
\newcommand{\ServTime}{X}
\newcommand{\TaskSize}{Y}
\newcommand{\SchedPolicy}{\pi}

%\newpage

\title{Synergy via Redundancy: Adaptive Replication Strategies and Fundamental Limits}
%\title{Boosting the Throughput of a Multi-server system\\ via Adaptive Task Replication}
%\title{Whole is Greater than the Sum of Parts: Boosting Server Throughput with Adaptive Replication}
%\title{Boosting the Throughput of a Server Cluster\\ with Adaptive Replication}

%\author{
%Gauri Joshi\\
%IBM T. J. Watson Research Center\\
%Yorktown Heights NY 10598\\
%Email: gauri.joshi@ibm.com
%}

\author{Gauri~Joshi,~\IEEEmembership{Member,~IEEE,}
        and~Dhruva Kaushal,~\IEEEmembership{Student Member,~IEEE}% <-this % stops a space
\thanks{G. Joshi (gaurij@andrew.cmu.edu) and D. Kaushal (dkaushal@alumni.cmu.edu) are with the Department
of Electrical and Computer Engineering, Carnegie Mellon University, Pittsburgh PA 15213}% <-this % stops a space
\thanks{A short 3-page abstract of this paper appeared in the 2017 ACM Sigmetrics MAMA Workshop, and a 6-page version appeared in IFIP WG 7.3 Performance 2018 \cite{joshi2018synergy}.}
\thanks{Manuscript accepted for publication in the IEEE/ACM Transactions on Networking.}
}


\markboth{}%IEEE/ACM Transactions on Networking,~Vol.~X, No.~Y, Dec~2020}%
{Shell \MakeLowercase{\textit{et al.}}: Bare Demo of IEEEtran.cls for IEEE Journals}
%
%
%\author{
%\IEEEauthorblockN{Gauri Joshi}
%\IEEEauthorblockA{Dept. 
%of ECE \\
%Carnegie Mellon University \\
%Pittsburgh, PA 15213\\
%Email: gaurij@andrew.cmu.edu}
%\and
%\IEEEauthorblockN{Dhruva Kaushal}
%\IEEEauthorblockA{Dept. 
%of ECE \\
%Carnegie Mellon University \\
%Pittsburgh, PA 15213\\
%Email: dkaushal@andrew.cmu.edu}
%}
%
%
%\author{
%Gauri Joshi \\
%Electrical and Computer Engg.\\
%Carnegie Mellon University \\
%Pittsburgh PA 15213\\
%Email: gaurij@andrew.cmu.edu
%}

\maketitle

%\listoftodos

\begin{abstract}
The maximum possible throughput (or the rate of job completion) of a multi-server system is typically the sum of the service rates of individual servers. Recent work shows that launching multiple replicas of a job and canceling them as soon as one copy finishes can boost the throughput, especially when the service time distribution has high variability. This means that redundancy can, in fact, create synergy among servers such that their overall throughput is greater than the sum of individual servers. This work seeks to find the fundamental limit of the throughput boost achieved by job replication and the optimal replication policy to achieve it. While most previous works consider upfront replication policies, we expand the set of possible policies to delayed launch of replicas. The search for the optimal adaptive replication policy can be formulated as a Markov Decision Process, using which we propose two myopic replication policies, MaxRate and AdaRep, to adaptively replicate jobs. In order to quantify the optimality gap of these and other policies, we derive upper bounds on the service capacity, which provide fundamental limits on the throughput of queueing systems with redundancy.
\end{abstract}

\section{Introduction}
\label{sec:intro}

%\subsection{Motivation and Overview}
\IEEEPARstart{T}{he} emergence of cloud computing services allows users who rent servers from service providers such as Amazon, Microsoft, and Google to seamlessly scale up or scale down their computational resource usage as per user demand. In order to offer this scalability and flexibility at extremely low cost, cloud service providers employs large-scale sharing of resources. Each server is shared by multiple users as well as background processes in both time and computing bandwidth. Such resource sharing is not centrally coordinated but rather the result of several schedulers operating independently. An adverse effect of large-scale resource sharing in cloud computing systems is that the response time of individual servers can be large and unpredictable. This inherent variability in response time is the norm and not an exception \cite{dean2013tail}. A simple yet powerful solution to combat service time variability is to replicate computing jobs at multiple servers and wait for any one copy to finish. This idea was first used at a large-scale in MapReduce \cite{dean2008mapreduce} and further developed in several other systems works including \cite{zaharia_spark_2010, zaharia_sparrow_2013}. A similar idea has been previously studied in \cite{maxemchuk2} to route packets in networks and in \cite{vulimiri2013low} in the context of DNS queries. 

\begin{figure}[t]
\centering
 \includegraphics[width= 2.0in]{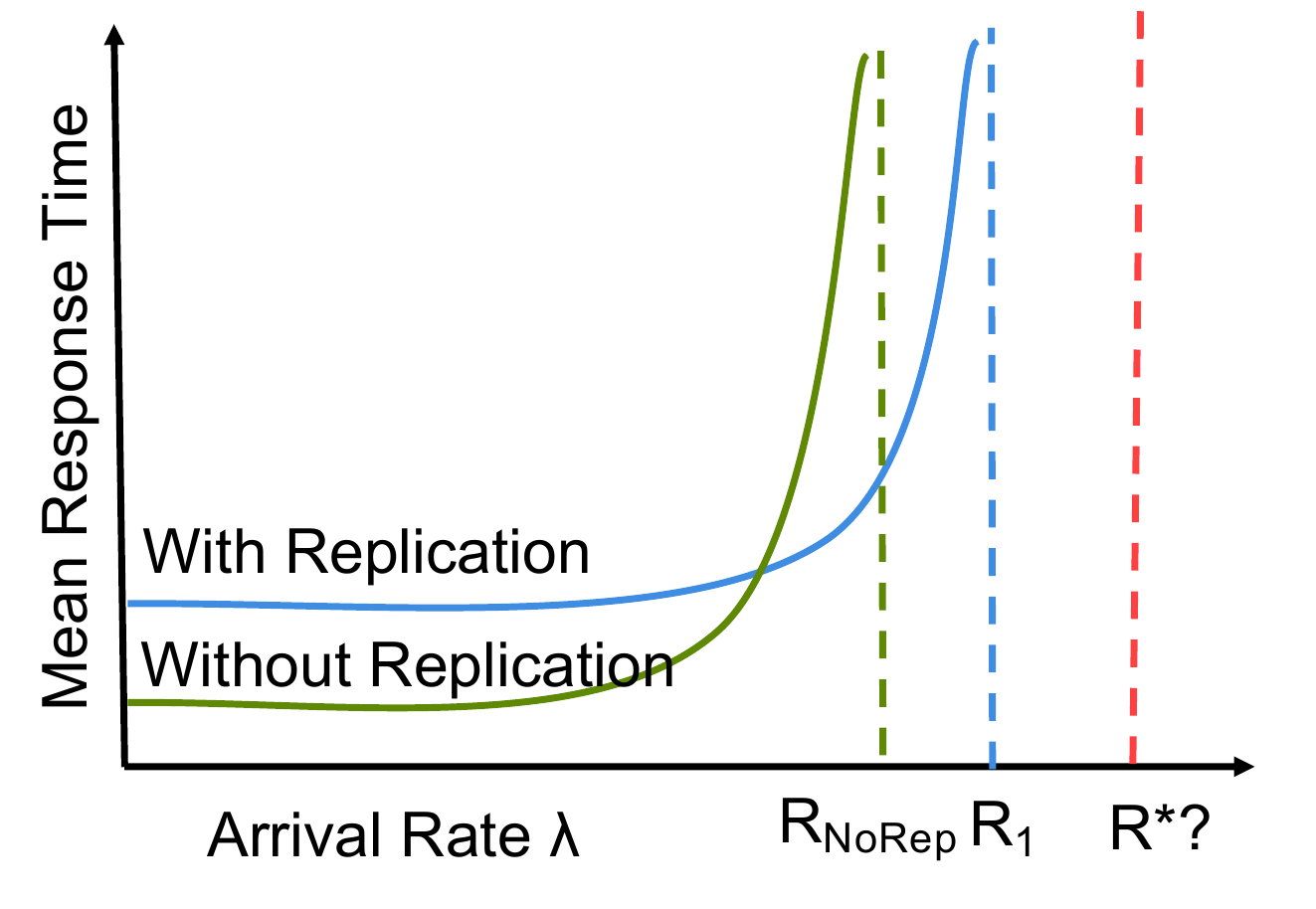}
\caption{All scheduling policies with replication achieve a maximum possible throughput $R_{No Rep}$, which is the number of the service rates of the servers in the systems. When service times have high variability,  replicating jobs at multiple servers and canceling replicas as soon as one copy finishes can yield a throughput that is greater than $R_{No Rep}$. Our goal is to find the fundamental limits of this throughput-boost and the replication policy that achieves it. \label{fig:serv_capacity}}
\vspace{-0.25cm}
\end{figure}

Although job replication is used in practical systems, only a few theoretical works provide an understanding of when redundancy is most beneficial in reducing latency. Works such as \cite{koole2008resource, joshi2014delay, shah2016when, sun2015provably, gardner2015reducing, joshi2015efficient} analyze multi-server queueing systems with redundancy. In these works, incoming jobs are replicated upfront and join queues at multiple servers simultaneously. As soon as any one replica finishes, all its copies are canceled immediately. Job replication affects response time (waiting time in queue plus service time) in two opposing ways:
\begin{enumerate}
\item \textit{Queue Diversity}: Replicas provide diversity by help finding the shortest among the queues that they join, thus reducing the overall waiting time. Unlike the join-the-shortest queue or power-of-choice scheduling policies which consider the queue lengths without accounting for the service time realizations of queued jobs, replication allow us to find queues that will be the first to become idle. This effect was studied in \cite{joshi2017efficient, gardner2015reducing}. 
\item \textit{Load due to Redundant Service}: A downside of launching more replicas is that more than one replicas may enter service at different servers, potentially adding load to the system and increasing the waiting time for subsequent jobs. However, \cite{koole2008resource, shah2016when, joshi2012coding} identify surprising scenarios where replicating jobs (and canceling the copies as soon as any one finishes) can in fact reduce the system load, and result in the overall system throughput being higher than the sum of service rates of individual servers.
\end{enumerate}
This paper seeks to dive deeper into the second factor, the effect of redundancy on the system load and throughput. We seek to find replication policies that achieve the maximum achievable throughput, which is also referred to as the \emph{service capacity}, denoted by $R^*$ in \Cref{fig:serv_capacity}. To the best of our knowledge, this is the first paper to attempt finding the service capacity with adaptive (rather than upfront) replication of jobs. Our system model accounts for server heterogeneity, job size variability as well as delays in cancellation of replicas. %We consider a large class of replication policies including upfront replication (launching all replicas at the same time) as well as adaptive replication, where replicas are added gradually, only if the original job does not finish in reasonable time. 

%Finding the optimal adaptive policy involves solving a Markov Decision Process (MDP). We formulate this MDP in \Cref{sec:mdp_formu}. This MDP can have a large state space and we need to resort to myopic policies. We propose two policies: MaxRate and AdaRep that adaptively add replicas and perform better than upfront replication policies. To quantify the gap from optimality we give an upper bound on the service capacity for the two-server case. All proofs are deferred to the Appendix, which can be found in the extended version \cite{hetero_servers_extended}.

\subsection{Related Work}

%\TODO{Cite more papers} 
%\TODO{Ciucu, Parag papers}

\textbf{Job Dispatching Policies in Multi-server systems.} Job dispatching and scheduling policies for multi-server systems have been studied for many decades in queueing theory \cite{kleinrock1975theory}. Traditionally metrics such as throughput and mean response time were studied in the context of operations and manufacturing systems. Queueing theory re-emerged in the early nineties as a rigorous way to design and analyze scheduling algorithms for computer systems \cite{mor_book}. However, \emph{queues with redundancy were not considered in queueing theory until recently because in operations research and early computer systems, the variability in the service time was largely due to randomness in job sizes}. Thus, replicating jobs offered no benefit in terms of reducing latency and thus the maximum possible throughput or the service capacity of such multi-server systems was simply the sum of the service rates of its servers. There is a rich line of literature on designing and analyzing the mean response time of job dispatching policies such as join-the-shortest-queue, power-of-$d$ choices \cite{powerof2}, least-work-left policy etc., proving their throughput-optimality and analyzing their mean response times. While the choice of the dispatch policy affects the mean response time, it does not change the maximum achievable throughput or the service capacity of the system. 

\vspace{0.15cm} \noindent \textbf{Redundancy to Overcome Delay Variability.} In the early 2000's, computing began to shift from local servers to the cloud, where computing resources are shared at a massive scale with limited central co-ordination. Although such loosely coordinated resource-sharing provides tremendous benefits in terms of cost, flexibility, and scalability, it causes \emph{random fluctuations in the server response times}. This service time variability is often referred to as ``tail latency". Due to tail latency, the same job can take vastly different execution times at two different servers \cite{dean2013tail}. The adverse effect of tail latency is further magnified in jobs with many parallel tasks because the probability of at least one of the tasks being a straggler increases exponentially. To overcome stragglers, several heuristic redundancy approaches such as back-up tasks, clones or hedged requests \cite{dean2008mapreduce, ananthanarayanan2013effective} began to be employed in computer systems. Although frequently used in systems, the addition of redundancy to overcome service time variability is a new paradigm in queueing theory with little understanding its fundamental limits. 

\vspace{0.15cm} \noindent \textbf{Fundamental Limits of the ``Free Lunch" Offered by Redundancy.} Only a few theoretical works \cite{koole2008resource, gardner2015reducing, gardner2016power, gardner2016s&x, joshi2015queues, shah2016when, joshi2017efficient, ayesta2018unifying, raaijmakers2019delta} have rigorously studied the latency of queues with redundancy and proposed redundancy scheduling policies. These works demonstrate that redundant jobs are extremely effective in finding the shorter queues in the system. However, intuition suggests that this benefit comes at the cost of additional load to system. This is because, when two or more replicas enter service, they use additional and redundant computing time of the servers and cause subsequent jobs to wait longer in queue. Contrary to this intuition, recent works \cite{koole2008resource, shah2016when, joshi2017efficient, joshi2018synergy, wang2019efficient, aktas2017service, aktas2020service, poloczek2016contrasting} identify regimes (when service times have high delay variability) where\emph{ redundancy can not only reduce overall latency but boost the throughput of queueing systems}. In \cite{anton2019stability} the authors analyze the increase in the service capapity (or the stability region) with different redundancy dispatch policies. \cite{raaijmakers2019redundancy} analyzes the service capacity for scaled Bernoulli service times whereas \cite{anton2020improving} analyzes the service capacity of processor sharing systems with heterogeneous service rates. However, these works only consider \emph{upfront replication policies} where all the replicas at launched at the same time -- delayed launch of replicas depending on the elapsed time of the original job has only been studied in  \cite{wang2019efficient} for parallel computing tasks without considering the effect of queueing of jobs. Understanding the fundamental limits of the ``free lunch" offered by redundancy is a unique and unexplored problem in queueing theory. And designing optimal redundancy strategies to take full advantage of this free lunch is of critical importance since it can help boost the efficiency of data centers and reduce their energy consumption.

\vspace{0.15cm} \noindent \textbf{Replication and Erasure Coding in Jobs with many parallel tasks.} Erasure codes, originally designed for error-correction and reliable transmission of information over a lossy communication channel, are a generalization of replication. Beyond their error-correction application, erasure codes can also be used to reduce delay and overcome stragglers in jobs with a large number of parallel tasks. For example, \cite{joshi2014delay, shah2016when} considered the problem of reducing the download time of a content file that is divided into $k$ chunks and coded into $n$ chunks using a maximum-distance-separable (MDS) code. Erasure coding allows us to recover the file from any $k$ out of $n$ chunks. Recently, erasure codes have also been shown to be effective in mitigating stragglers in parallel computing tasks such as matrix computations and distributed inference \cite{lee2017speeding, yu2017polynomial, dutta2016short, kosaian2019parity, mallick2019rateless, mallick2020rateless}. Analyzing the mean response time experienced by such jobs with $n$ parallel tasks where it is sufficient to complete any $k$ tasks is equivalent to an $(n,k)$ fork-join system. It is a generalization of the fork-join queueing system \cite{varki_merc_chen, nelson_tantawi, rizk_poloczek_ciucu_2015}, which is a notoriously  hard problem even for exponential service times. Papers such as \cite{joshi2014delay, shah2016when, xiang2016joint, parag2017latency, badita2019latency} give bounds on the latency of the $(n,k)$ fork-join system while others such as \cite{li_mean_field_2016} use mean-field analysis to compare replication and erasure coding. Instead of latency, in this paper, we focus on the maximum achievable throughput or the service capacity with job replication. Going beyond replication, characterizing the service capacity of erasure-coded storage and computing systems is an open future problem and has been considered in only a few recent works \cite{aktas2017service, aktas2020service}.

\vspace{0.15cm} \noindent \textbf{Main Differences from Previous Works.} To summarize, the problem formulation of this paper differs from prior works on redundancy in queueing systems in the three key ways: 1) considering non-exponential service times for which the service capacity can potentially be increased using job replication, 2) instead of upfront replication, we consider gradual launch of additional replicas in order to preserve high throughput, and 3) the first attempt (to the best of our knowledge) to determine the service capacity, that is, the maximum possible throughput of a multi-server system with job replication under these general conditions. The replication strategies proposed in this paper are analyzed in terms of their throughput-optimality. Since throughput-optimality of replication strategies is still not well-understood, the much harder problem of designing of delay-optimal policies that minimize the mean response time for any given arrival rate $\lambda$ is beyond the scope of this work. However, through simulations we show that the proposed replication strategies work well in the low and moderate traffic regimes.

\section{Problem Formulation}
\label{sec:prob_formu}

\begin{figure}[t]
\centering
 \includegraphics[width= 3.0 in]{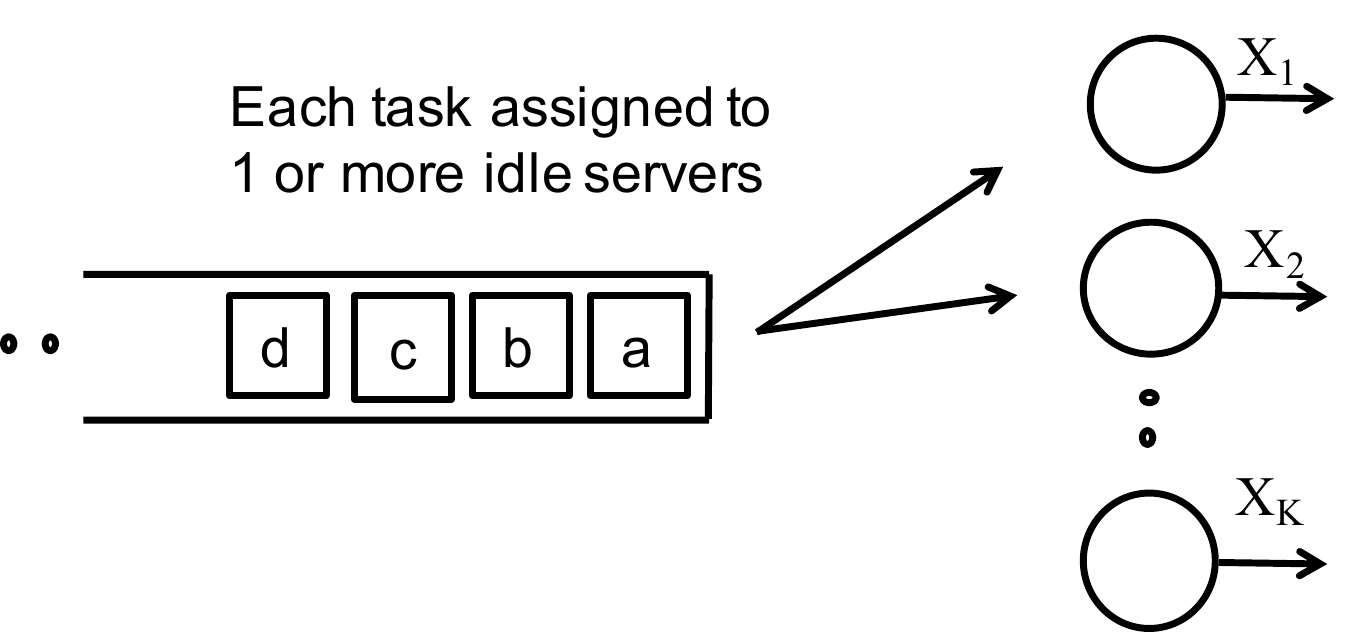}
\caption{System of $\NumServers$ servers where a job replicated at two idle servers $1$ and $2$ takes time $\min(X_1, X_2)$ to finish, where the random variable $X_i$ captures the service time variability.\label{fig:het_servers_sys_model}}
%\vspace{-0.25cm} 
\end{figure}

Consider a system of $\NumServers$ servers with a central queue containing jobs, illustrated in \Cref{fig:het_servers_sys_model}. These jobs are served in a first-come-first-served manner, and each job can be assigned to one or more idle servers. We do not explicitly define a job arrival process and instead assume that the central queue is never idle. Since our objective is to maximize the throughput, or the rate of job completion, there is no loss of generality in this assumption. Only in \Cref{sec:response_time}, we consider Poisson job arrivals with rate $\lambda$ into the central queue -- this is for the purpose of simuations that demonstrate the mean response times of the proposed replication policies in low or moderate traffic regimes. %assumption does not result in the loss owe do not explicitly consider the job arrival process and instead assume that the central queue is never idle. %Below we describe the service time distributions, performance metrics and scheduling policies considered in this paper.

\subsection{Job Service Times}
The system consists of $K$ heterogeneous servers, where server $i$ takes time $\ServTime_i$ to finish a job assigned to it, where the probability distribution of $X_i \sim F_{X_i}$. The random variable $X_i$ captures the variability in job service time due to server slowdown, assumed to be i.i.d.\ across jobs assigned to that server. \footnote{We assume that the service time variability comes only from the server and not the size of the job. However, it is possible to account for job size variability via another random variable $Y$, which is independent of $X_i$ for all $i$. This method of multiplying the randomness from the two sources of variability was introduced in \cite{gardner2016s&x}. The value of $Y$ is same across replicas of a job. Thus, if a job is replicated at two idle servers $i$ and $j$, the time taken to complete any one replica is $Y \cdot \min(X_i, X_j)$. For simplicity and brevity we assume $Y = 1$ (deterministic job size) in this paper, but most of the results can be extended to random $Y$ by adding a scaling factor $\E{Y}$ multiplying $\E{\min(X_i, X_j)}$. } We also consider that when a job is replicated, each server running it (including the server running the original copy of the job) reserves a cancellation window of length $\Delta$. As soon as one replica is served, the scheduler sends a cancellation signal to the other replicas, triggering their cancellation. All these events occur in time $\Delta$, after which the servers are available to serve subsequent jobs. 

\subsection{Scheduling Policy}
The policy $\pi$ used to schedule replicas can be based on the service time distributions of $X_1$, \dots,  $X_K$. The scheduler only knows these distributions, but does not know their realizations for currently running jobs. As soon as a server becomes idle, the scheduler can take one of two possible actions:
\begin{itemize}
\item \emph{\textbf{new}}: assign a new job to that server
\item \emph{\textbf{rep}}: launch a replica of a job currently running on one of the other servers.
\end{itemize}
The space of scheduling policies with these actions is denoted by $\Pi_{n,r}$ and we aim to find the policy $\pi^*_{n,r}$ that maximizes the throughput. This space of policies can be expanded by allowing additional actions such as pausing a currently running job, or killing and relaunching it to another server. We only focus on the \emph{\textbf{new}} and \emph{\textbf{rep}} actions in this paper. Only in \Cref{sec:converse_bnd} we use job pausing to find an upper bound on the service capacity.

Note that all job replication policies in $\Pi_{n,r}$ are work-conserving, that is, they do not allow any server to be idle for a non-zero time interval. \Cref{clm:work_conserving_optimality} below shows that there is no loss of generality in restricting our attention to work-conserving policies. 

\begin{clm}
\label{clm:work_conserving_optimality}
The throughput-optimal scheduling policy $\SchedPolicy^*$ is work-conserving, that is, it does not allow any server to be idle for a non-zero time interval.
\end{clm}

The proof is given in the Appendix.

\subsection{Performance Metrics}
Let us formally define the throughput of policy $\SchedPolicy$.

\begin{defn}[Throughput $\Thpt$]
\label{defn:throughput}
Let $T_1(\SchedPolicy) \leq T_2(\SchedPolicy)\leq \dots \leq T_n(\SchedPolicy)$ be the departure times of jobs $1, 2, \dots n$ from the system, when the scheduler follows a policy $\SchedPolicy$. Then the throughput is defined as
 \begin{align}
\Thpt(\SchedPolicy) \triangleq \lim_{n \rightarrow \infty} \frac{n}{T_n(\pi)}. \label{eqn:thpt_in_terms_of_total_delay}
\end{align}
\end{defn}

\begin{defn}[Service Capacity $R^*_{n,r}$]
The service capacity $R^*_{n,r} = \max_{\pi \in \Pi_{n,r}} R(\pi)$, the maximum achievable throughput over all scheduling policies in $\Pi_{n,r}$. The policy $\pi^*_{n,r}$ that achieves $R^*_{n,r}$ is called the throughput-optimal policy.
\end{defn}

An alternate interpretation of $R(\pi)$ is that if jobs are arriving in the central queue at rate $\lambda$, then if $\lambda < R(\pi)$ the system is stable, that is, the mean response time (waiting time plus service time) experienced by jobs is finite. Thus by using the throughput-optimal policy $\pi$ that maximizes $R(\pi)$, we can support the maximum possible job arrival rate.

Next we define another performance metric, the computing time $C$ per job. In \label{clm:thpt_in_terms_of_EC} we will show how throughput $R$ can be expressed in terms of $\E{C}$.

\begin{defn}[Computing Time $\Cost$]
\label{defn:computing_cost}
The computing time $\Cost$ is the total time collectively spent by the servers per job. 
\end{defn}

The expected computing time $\E{C}$ is proportional to the cost of running a job on a system of servers, for instance, servers rented from Amazon Web Services (AWS), which are charged by the hour. In our system model, if a job is assigned to only to server $i$ then $\E{C} = \E{X_i}$. Instead, if it is assigned to two servers $i$ and $j$, and the replica is canceled when any one copy finishes then $\E{C} = 2 (\E{\min(X_i, X_j)} + \Delta)$ where $\Delta$ is the cancellation window at each of the servers. Depending upon $X_i$ and $\Delta$, $\E{C}$ with replication may be greater or less than that without replication. 

\begin{clm}
\label{clm:thpt_in_terms_of_EC}
For any work-conserving scheduling policy, 
\begin{align}
\Thpt = \frac{\NumServers}{\E{\Cost}}. \label{eqn:thpt_in_terms_of_C}
\end{align}
\end{clm}

\begin{proof}
Consider jobs $1$, $2$, \dots $n$ run on the system of servers. If the scheduling policy is work-conserving, the total busy time of each server is exactly equal to $T_n$, the departure time of the last job. Since $\E{C}$ is defined as the total expected time spent at servers per job, by law of large numbers we have
\begin{align}
 \E{\Cost} &= \lim_{n \rightarrow \infty} \frac{K T_n}{n} = \frac{K}{\Thpt},
\end{align}
where the second equality follows from \Cref{defn:throughput}.
%Substituting this and the definition of $\E{C}$,
%\begin{align}
%\Thpt = \lim_{n \rightarrow \infty} \frac{T_n}{n} =  \frac{\E{\Cost}}{\NumServers}
%\end{align}
\end{proof}

Thus, minimizing $\E{\Cost}$ is equivalent to maximizing $\Thpt$. 
%\TODO{The work-conserving claim should appear before we define the space of scheduling policies under consideration in this paper}

\subsection{Main Contributions and Organization}
\label{subsec:main_ideas}
To illustrate the main contributions and organization of this paper, let us consider some replication policies for a simple two-server example. The upcoming sections will develop each of the replication policies considered in this example in greater detail and rigor.

\begin{figure}[t]
\centering
 \includegraphics[width= 3.5 in]{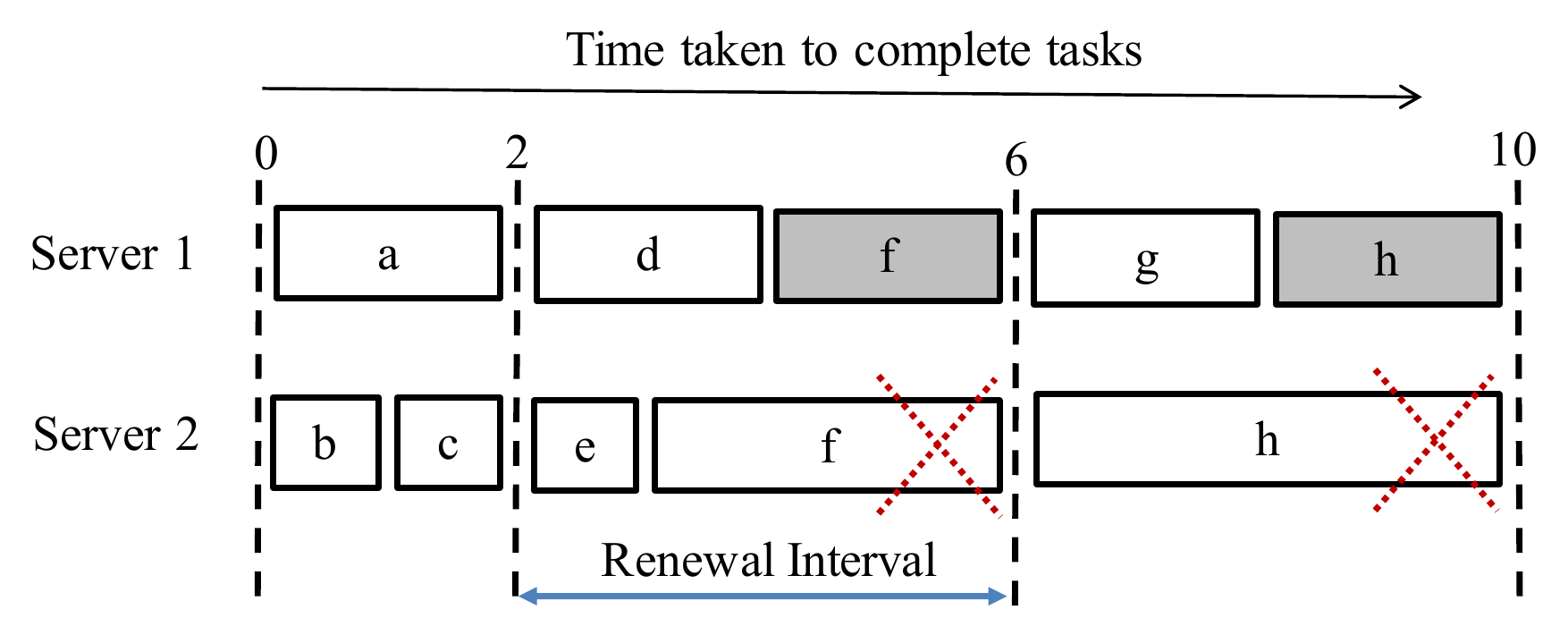}
\caption{Illustration of renewal instants of the system of $2$ servers with the adaptive replication policy ($\pi_{AdaRep}$) described in \Cref{subsec:main_ideas}.\label{fig:adarep_exmple}}
\end{figure}

\begin{exple}
\label{exple:adarep_exmple}
Consider a sytem of two servers with service time distributions
\begin{align}
\ServTime_1 &= 2 \\
\ServTime_2 &= 
\begin{cases}
1  &\text{w.p.} \quad  (1-p)=0.9 \\
20 &\text{w.p.} \quad p = 0.1
\end{cases}
\end{align}
The cancellation delay $\Delta = 0$. The throughput or the rate of job completion with full replication and no replication respectively are
\begin{align}
\Thpt_{NoRep} &= \frac{1}{\E{\ServTime_1}} + \frac{1}{\E{\ServTime_2}} = 0.8448, \\
\Thpt_{FullRep} &= \frac{1}{\E{\min(X_1, X_2)}}  = 0.909.
\end{align}

In \Cref{sec:extreme_policies} we analyze more general `upfront' replication policies that launch $ r < K$ replicas of a job at the same time and cancel the outstanding replicas as soon as any one copy is served. An alternative to upfront replication is to add replicas gradually, only if the original copy of the job does not finish in reasonable time. One such policy is the adaptive replication (AdaRep) policy $\pi_{AdaRep}$, which launches a replica of a job assigned to server $2$ only if it has spent more than $1$ second in service. To evaluate the throughput of this policy, we consider time instants called renewals when both servers become idle. There are three types of intervals between successive renewal instants as illustrated in \Cref{fig:adarep_exmple}. The throughput is the expected number of jobs completed in an interval, divided by the expected interval length.
\begin{align}
\Thpt_{AdaRep} 
&=  \frac{\sum_{i = 1}^{3} \Pr( \text{Type i interval}) \cdot (\# \text{jobs completed})}{ \text{Expected length of a renewal interval}}\\
&= \frac{0.9 \times 0.9 \times 3 + 0.9 \times 0.1 \times 3 + 0.1 \times 2 }{0.9 \times 0.9 \times 2 + 0.9 \times 0.1 \times 4 + 0.1 \times 4}\\
&\approx 1.2185,
\end{align}
which clearly outperforms the two extreme policies.
\end{exple}

For systems with more general service distributions, finding the optimal adaptive policy involves solving a Markov Decision Process (MDP). We formulate this MDP in \Cref{sec:mdp_formu}. This MDP can have a large state space and we need to resort to myopic policies. We propose two such policies -- MaxRate and AdaRep in \Cref{sec:myopic_policies}. The MaxRate policy which launches replicas so as to maximize the instantaneous job departure rate from the system. The AdaRep policy that launches replicas when the elapsed time currently running copies crosses a pre-specified threshold. In order to quantify the optimality gap of these policies, in \Cref{sec:converse_bnd} we obtain upper bounds on the service capacity for the two heterogeneous servers case, and a more general upper bound for $K$ homogeneous servers. Finally \Cref{sec:conclu} presents major implications and future directions. %And the upper bound inspires an adaptive replication policy.

\subsection{Notations used in the paper}
In the rest of the paper we use $[K]$ to denote the set $\{1, 2, \dots K\}$. The symbol $X_{k:n}$ is used to denote the $k$-th order statistic (or the $k$-th smallest) of $n$ i.i.d. realizations $X_1, X_2, \dots X_n$ of the random variable $X$. For example, $X_{1:r} = \min(X_1, X_2, \dots X_r)$, the minimum of $r$ i.i.d.\ realizations of $X$. 

Finally, the residual part of a non-negative random variable $X \geq 0$, given that $X>t$ is denoted by $X^{rs}(t) = (X - t)|X > t$, which has the complementary cumulative (or tail probability) distribution function $Pr(X^{rs}(t) > x) =  \Pr(X >t+x)/\Pr(X > t)$. We also define the truncated version of $X$ as $X^{tr}(t) = \min(X, t)$.

%\textbf{Job-shop Scheduling}
%\textbf{Flexible server systems}
%\textbf{Replication in Storage and Computing}
%\textbf{Exponential Service Time}
%\textbf{General Service Time}
%
%\textbf{Heterogeneous servers}
%
%
%In this work our aim is to answer two main questions that arises from the discussion above:
%\begin{itemize}
%\item What is the maximum throughput we can get from a system of servers when job replication is allowed? In other words, what is the service capacity?
%\item How can we provide fair service to jobs when the servers are heterogeneous? 
%\end{itemize}

%The rest of the paper is organized as follows. \Cref{sec:extreme_policies} compares two extreme policies: no replication and full replication and provides insights into when one outperforms the others. A softer replication policy can outperform these two extremes. In \Cref{sec:mdp_formu} we provide a Markov Decision Process (MDP) framework to find the optimal policy, which can outperform the two extreme policies. This MDP is hard to solve in general and myopic policies are not close to optimal. In \Cref{sec:converse_bnd} we give an upper bound on the service capacity for the two heterogeneous servers case and a more general upper bound for $K$ homogeneous servers. To find the upper bound, we employ a novel genie system called the pause-and-replicate system. Finally \Cref{sec:conclu} presents major implications and future directions.

\section{Upfront Replication}
\label{sec:extreme_policies}

In this section we explore `upfront' replication policies that simultaneously launch a job and its replicas. The number of replicas and the servers where they are launched governs the overall throughput. 

\subsection{No Replication and Full Replication}
First let us compare the throughput achieved by two extreme policies: no replication and full replication. This analysis demonstrates how replication can create synergy and boost the throughput of a server cluster. 

%\begin{figure}[t]
%\centering
%\begin{subfigure}[t]{0.85\linewidth}
%    \centering
%   \includegraphics[width=2.5in]{no_rep_model.pdf}
%	\caption{No replication ($\SchedPolicy_{NoRep}$): Assign each job to the earliest available idle server}
%\end{subfigure}

%\begin{subfigure}[t]{0.85 \linewidth}
%    \centering
%   \includegraphics[width=2.5in]{full_rep_model.pdf}
%\caption{Full replication ($\SchedPolicy_{FullRep}$): Replicate each job at all servers. As soon as one replica finishes, cancel the rest}
%\end{subfigure}
%\caption{Illustration of the no replication and full replication policies.\label{fig:het_two_servers}}
%\end{figure}
%
%\begin{figure*}[t]
%    \centering
%    \begin{subfigure}[b]{0.31\textwidth}
%        \centering
%        \includegraphics[width =6.3cm]{full_rep_no_rep_hyper_exp_exp}
%        \caption{}
%    \end{subfigure}
%    \begin{subfigure}[b]{0.31\textwidth}
%        \centering
%        \includegraphics[width = 6.3cm]{full_rep_no_rep_shifted_exp_pareto}
%        \caption{}
%    \end{subfigure}
%    \begin{subfigure}[b]{0.31\textwidth}
%        \centering
%        \includegraphics[width = 6.3cm]{full_rep_no_rep_shifted_exp_shifted_hyperexp}
%        \caption{}
%    \end{subfigure}
%    \caption{Caption place holder}
%\end{figure*}

\begin{lem}[Throughput with No Replication]
\label{lem:thpt_no_rep}
If each job is assigned to the first available idle server in a system of $K$ servers, the throughput is,
\begin{align}
\Thpt_{NoRep} &= \sum_{i=1}^{K} \frac{1}{\E{X_i}} \label{eqn:thpt_norep}
\end{align}
\end{lem}

\begin{proof}
This policy is work-conserving and thus keeps all servers busy all the time. Thus, if we look at server $i$, the departure time of the $n^{th}$ job assigned to that server is $T_n^{(i)}$ is the sum of $n$ i.i.d. realizations of the service time $X_i$. Thus, the rate of departure of jobs from server $i$ is,
\begin{align}
\Thpt_i &= \lim_{n \rightarrow \infty} \frac{ n}{T_n^{(i)}} = \frac{1}{\E{X_i}}.
\end{align}
Adding the rates of departure from all the servers yields overall throughput as given by \eqref{eqn:thpt_norep}.
\end{proof}

\begin{lem}[Throughput with Full Replication]
\label{lem:thpt_full_rep}
Suppose each job is assigned to all servers, and as soon as one replica finishes, the others are canceled. The throughput achieved by this full replication policy is,
\begin{align}
\Thpt_{FullRep} &= \frac{1}{\Delta +\E{\min(X_1, X_2, \dots X_K)}} \label{eqn:thpt_fullrep}
\end{align}
\end{lem}

\begin{proof}
With the full replication policy, all $K$ servers are working on the same job at any time instant. The total time spent by them on each job is,
\begin{align}
\E{C} &= K (\Delta + \E{\min(X_1, X_2, \dots X_K)})
\end{align}
Then \eqref{eqn:thpt_fullrep} follows from the result in \Cref{clm:thpt_in_terms_of_EC}.
\end{proof}

% \begin{figure}[t]
%    \centering
%   \includegraphics[width= 3.5 in]{het_two_serv_log_concave_convex.pdf}
%\caption{Comparison of the no replication and full replication policies for $X_1 \sim \Exp(0.5)$ and different $X_2$. When $X_2 \sim \SExp(c,0.25)$, a shifted exponential no replication is always better (illustrated on the left). When $X_2 \sim \HyperExp(p=0.3, \mu_1 = 0.5, \mu_2)$, a hyper-exponential then full replication is better (illustrated on the right).
% \label{fig:het_two_serv_log_concave_convex}}
%\end{figure}

%%\subsubsection{$\SchedPolicy_1$ versus $\SchedPolicy_2$ comparison}
%\begin{coro}
%\label{lem:two_server_opt_policy}
%Consider the two server case, and assume that $Y= 1$. Then the full replication gives higher throughput than no replication if and only if
%\begin{align}
%\frac{1}{\Delta + \E{\min(X_1, X_2)}} > \frac{1}{\E{X_1}} + \frac{1}{\E{X_2}} \label{eqn:two_server_opt_policy}
%\end{align}
%\end{coro}

\begin{figure}[t]
    \centering
   \includegraphics[width= 3.5 in]{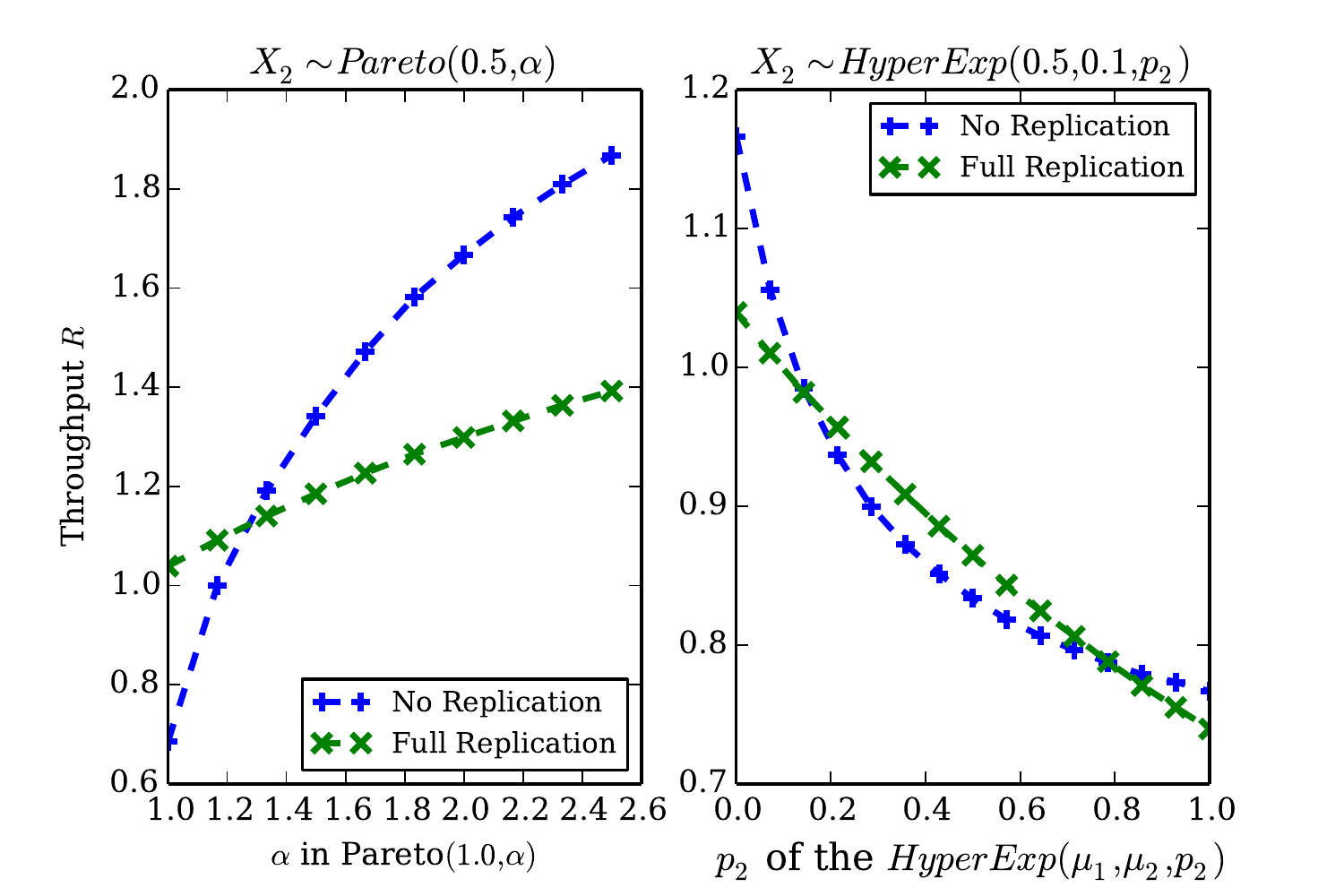}
\caption{Comparison of the no replication and full replication policies for $X_1 \sim 0.5 + \Exp(1)$ and different $X_2$. When $X_2 \sim \Pareto(0.5,\alpha)$, full replication is better for smaller $\alpha$ (heavier tail). When $X_2 \sim \HyperExp(\mu_1 = 0.5, \mu_2=0.1, p_2)$, full replication is better for intermediate $p_2$.
 \label{fig:het_two_serv_log_concave_convex}}
\end{figure}

Using \Cref{lem:thpt_no_rep} and \Cref{lem:thpt_full_rep} we can compare the two policies for any given distributions $X_1, \dots, X_K$ and cancellation delay $\Delta$. In \Cref{fig:het_two_serv_log_concave_convex} we show a comparison of full replication and no replication for the two server case, with $\Delta = 0$. In both subplots, the service time $X_1\sim 0.5 + \Exp(1)$, a shifted exponential. We observe that full replication gives higher throughput when $X_2$ has higher variability. In the left subplot, $X_2 \sim \Pareto(0.5, \alpha)$ and replication is better for smaller $\alpha$ (heavier tail). In the right subplot, $X_2$ is a hyper-exponential $\HyperExp(\mu_1, \mu_2, p_2)$, that is, it is an exponential with rate $\mu_2$ with probability $p_2$ and otherwise it is exponential with rate $\mu_1$.  In this case, replication is better for intermediate $p_2$  where $X_2$ has higher variability. 

%From \eqref{eqn:two_server_opt_policy} we can also see that as the cancellation delay $\Delta$, the throughput of the full replication policy becomes worse. Thus if $\Delta$ exceeds some threshold, the no replication policy would outperform full replication.

\subsection{General Upfront Replication}
Instead of replicating job at all servers, or not replicating at all we can replicate jobs at a subset of the servers. Each subset is treated as a `super-server' such that jobs are replicated at all servers in a super-server. We refer to this class of policies as upfront replication policies, defined formally below.

\begin{defn}[Upfront Replication]
For positive integers $h \in \mathbb{N}$, consider a partition of set $[K] = {1, 2, 3, \dots K}$. The partition is a collection of non-empty subsets $\mathcal{S}_1, \mathcal{S}_2, \dots, \mathcal{S}_h$ of $[K]$, such that $\mathcal{S}_i \cap \mathcal{S}_j = 0$, and $\cup_{j} S_j = [K]$. When the servers in a set $\mathcal{S}_j$ become idle (they will always become idle simultaneously), assign the next job in the central queue to them.
\end{defn}

The no replication policy is a special case with $\mathcal{S}_j = \{j \}$ for all $j \in [K]$. Full replication is also a special case with $\mathcal{S}_1 = [K]$.

\begin{thm}[Throughput with Upfront Replication]
\label{thm:upfront_rep}
The throughput $R_{UpFr}$ achieved by upfront replication at server sets $\mathcal{S}_1, \dots, \mathcal{S}_h$ is
\begin{align}
&R_{UpFr}(\mathcal{S}_1, \dots, \mathcal{S}_h) = \sum_{j=1}^{h} \frac{1}{\E{X_{\mathcal{S}_j}} + \Delta}, \label{eqn:thpt_upfront_rep} \\
& \text{where } X_{\mathcal{S}_j} = \min_{l \in \mathcal{S}_j} X_l 
\end{align}
\end{thm}

The proof is given in the Appendix. To maximize the throughput, we need to find the partition $\{ \mathcal{S}_1, \dots, \mathcal{S}_h \}$ that maximizes \eqref{eqn:thpt_upfront_rep}. The number of possible partitions of a set of size $K$ is given by the Bell number $B_k$. It can be computed using the recursion
\begin{align}
B_K =\sum _{i=0}^{K-1}\binom {K-1}{i} B_{i},
\end{align}
with base $B_0 = 1$. This number is exponential in $K$. Thus, when the number of servers $K$, searching over all possible partitions to find the partition that maximizes the throughput can be computationally intractable. 

However, most practical multi-server systems consist of only a few types of servers, such that servers of the same type have the same service time distribution. Finding the best partition of the servers can be tractable in such systems. For example, for $K$ homogeneous servers with service time distribution $F_X$, the throughput of the optimal upfront replication policy is given by the following result. 

\begin{thm}[Upfront Replication Throughput Bound for $K$ Homogeneous Servers]
\label{thm:scaling_upfront_K_servers}
For a system of $K$ homogeneous servers with i.i.d. service times $X \sim F_X$, let $r^*$ be the positive integer that minimizes $r (\E{X_{1:r}} +\Delta)$. The throughput achieved with upfront replication of jobs satisfies
\begin{align}
R_{UpFr} &\leq \frac{K}{r^* (\E{X_{1:r}} + \Delta)}. \label{eqn:opt_thpt_upfront_rep}
\end{align}
Equality holds in \eqref{eqn:opt_thpt_upfront_rep} if $r^*$ divides the number of servers $K$. 
\end{thm}

%
%The $r^*$ that minimizes $r_i (\E{X_{1:r_i}} + \Delta)$ depends on the tail of the distribution. For $Y=1$ (no job size variability) and $\Delta = 0$, $r^*$ is the $r$ that minimizes $r \E{X_{1:r}}$. If the log of the tail distribution $\Pr(X>x)$ is concave (convex) in $x$, then the distribution is said to be log-concave (log-convex). For these two classes of distributions, we directly get the optimal $r^*$ by \Cref{lem:log_concave_convex} below. 
%
%\begin{lem}
%\label{lem:log_concave_convex}
%For a system of $K$ homogeneous servers, if the service time distribution $X$ is log-concave (log-convex), then $r^* = 1$ ($r^* = K$) minimizes $r \E{X_{1:r}}$.
%\end{lem}
%
%\begin{proof}
%It can be shown that if $X$ is log-concave (log-convex), then the function $r \E{X_{1:r}}$ is non-decreasing (non-increasing) in $r$, for positive integers $r$. The detailed proof is given in the Appendix.
%\end{proof}

\begin{figure}[t]
\centering
 \includegraphics[width= 3.5 in]{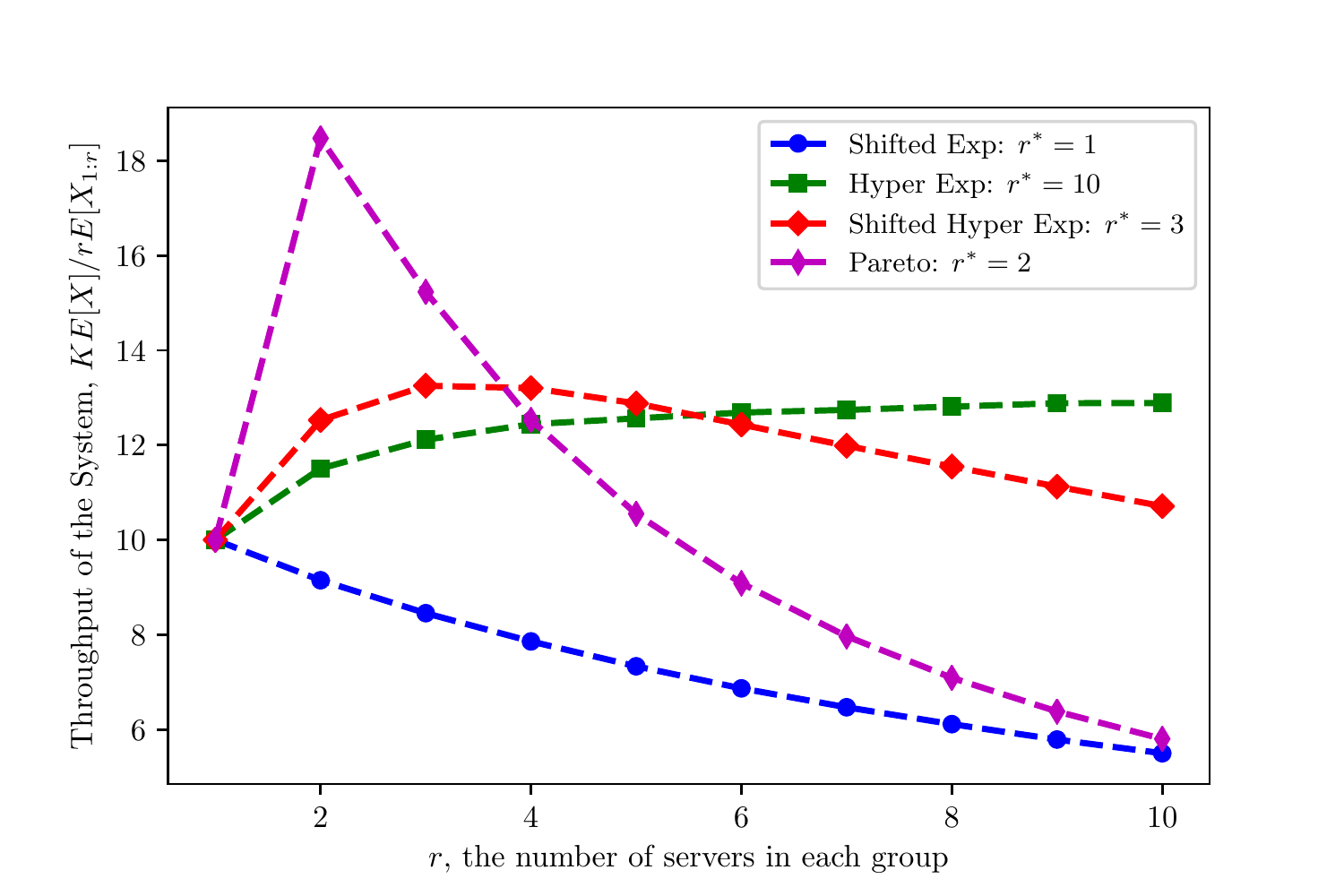}
\caption{The throughput of a system of $K=10$ servers versus the number of replicas $r$ per job for different service distributions, with $K = 10$. When $X$ is shifted-exponential (log-concave), the optimal $r$ is $r=1$, whereas for hyper-exponential $X$, $r^* = K$ is optimal. \label{fig:scaling_K_servers_throughput}}
\end{figure}

The proof is given in the Appendix.  For $\Delta = 0$, $r^*$ is the $r$ that minimizes $r \E{X_{1:r}}$. \Cref{fig:scaling_K_servers_throughput} illustrates the throughput of a system of $K$ servers, which is equal to $K \E{X}/r \E{X_{1:r}}$ versus $r$ for four different service distributions: shifted exponential $0.1 + \Exp(1.0)$, hyper-exponential $\HyperExp(0.6, 0.2, 0.4)$, shifted hyper-exponential $0.1 + \HyperExp(1.0, 0.2, 0.4)$, and Pareto $\Pareto(0.5, 1.2)$. When the tail distribution $\Pr(X>x)$ of  $X$ is log-concave (for example shifted-exponential), the optimal $r$ is $r=1$, whereas for log-convex $X$ (for example hyper-exponential), $r^* = K$ is optimal. This property of log-concave (log-convex) distributions was proved in \cite{joshi2017efficient}. For other distributions such as shifted hyperexponential or Pareto, intermediate $r$ can be optimal.  

\section{MDP Formulation of the Throughput-Optimal Replication Policy}
\label{sec:mdp_formu}

Instead of launching replicas upfront, they could be added conditionally, only if the original job does not finish in some given time. %Such policies can significantly increase the throughput, as illustrated by the example below.
We propose a Markov Decision Process (MDP) framework to search for the throughput-optimal policy the achieves service capacity. Finding the optimal policy directly using this framework is an intractable problem, but it provides valuable insights into the design of myopic policies in \Cref{sec:myopic_policies}. We describe the state-space, actions, and cost per transition below. Observe that state-space and actions satisfy the Markov property, that is, the transition from state $s$ to $s'$ only depends on the action $\SchedPolicy(s)$, and is conditionally independent of all previous states and actions.

\subsection{State-space}
We denote the state evolution by $s_0, s_1, \dots s_i, \dots$ such that the system transitions to state $s_i$ as soon as the $i^{th}$ job departs. The state-space can be collapsed into states $[\mathcal{B}, \mathbf{t}, D_r]$ where $\mathcal{B}$ contains disjoint sets of server indices that are running the unfinished jobs in the system and $|\mathcal{B}|$ is the number of jobs currently in the system. For example, if $\mathcal{B} = \{ \{1\}, \{2, 3\} \}$ there are $|\mathcal{B}| = 2$ unfinished jobs in the system, one running on server $1$ and another on servers $2$ and $3$. The vector $\mathbf{t} = (t_1, t_2, \dots t_K)$ where $t_k$ is the time spent by server $k$ on its current job. Since we observe the system immediately after a job departure, at least one of the elapsed times $t_1, t_2, \dots t_K$ is zero. The purpose of the $D_r$ term is to ensure that each state transition corresponds to a single job departure. It is the number of jobs that have finished, but are still to depart. If $h > 1$ jobs exit the system simultaneously and result in the job assignment set $\mathcal{B}$ and elapsed-time vector $\mathbf{t}$, then the system goes through states $[\mathcal{B} , \mathbf{t}, h-1] \rightarrow [\mathcal{B} , \mathbf{t},  h-2] \rightarrow \dots \rightarrow[\mathcal{B} , \mathbf{t}, 0]$. %The initial state $s_0 = [\emptyset, (0, \dots, 0), 0]$. 

%and result in given state-tr exit the system simultaneously, then we need to add exit states $E_1, \dots E_{i-1}$ to ensure that each transition corresponds to a single job departure. For example in \Cref{fig:adarep_exmple} the state is $(1,0)$ at time $t = 1$. At $t = 2$, two jobs depart simultaneously. We consider two state transitions occurring at this instants:  from $(1, 0)$ to $(0, 0, E_1)$, and then to $(0, 0)$. 

\subsection{Actions}
In each state $s$, denote the set of possible actions is $\mathcal{A}_s$. The scheduling policy $\SchedPolicy$ determines the action $a = \SchedPolicy(s)$ that is taken from state $s$. First note that no jobs are assigned in the exit states $s = [\mathcal{B}, \mathbf{t}, D_r]$ with $D_r > 0$. Thus, for these states, the action space $\mathcal{A}_s$ contains a single placeholder $\textbf{\emph{null}}$ action. The system directly transitions to $[\mathcal{B}, \mathbf{t}, D_r-1]$.

\begin{figure}[t]
\centering
 \includegraphics[width= 3.8 in]{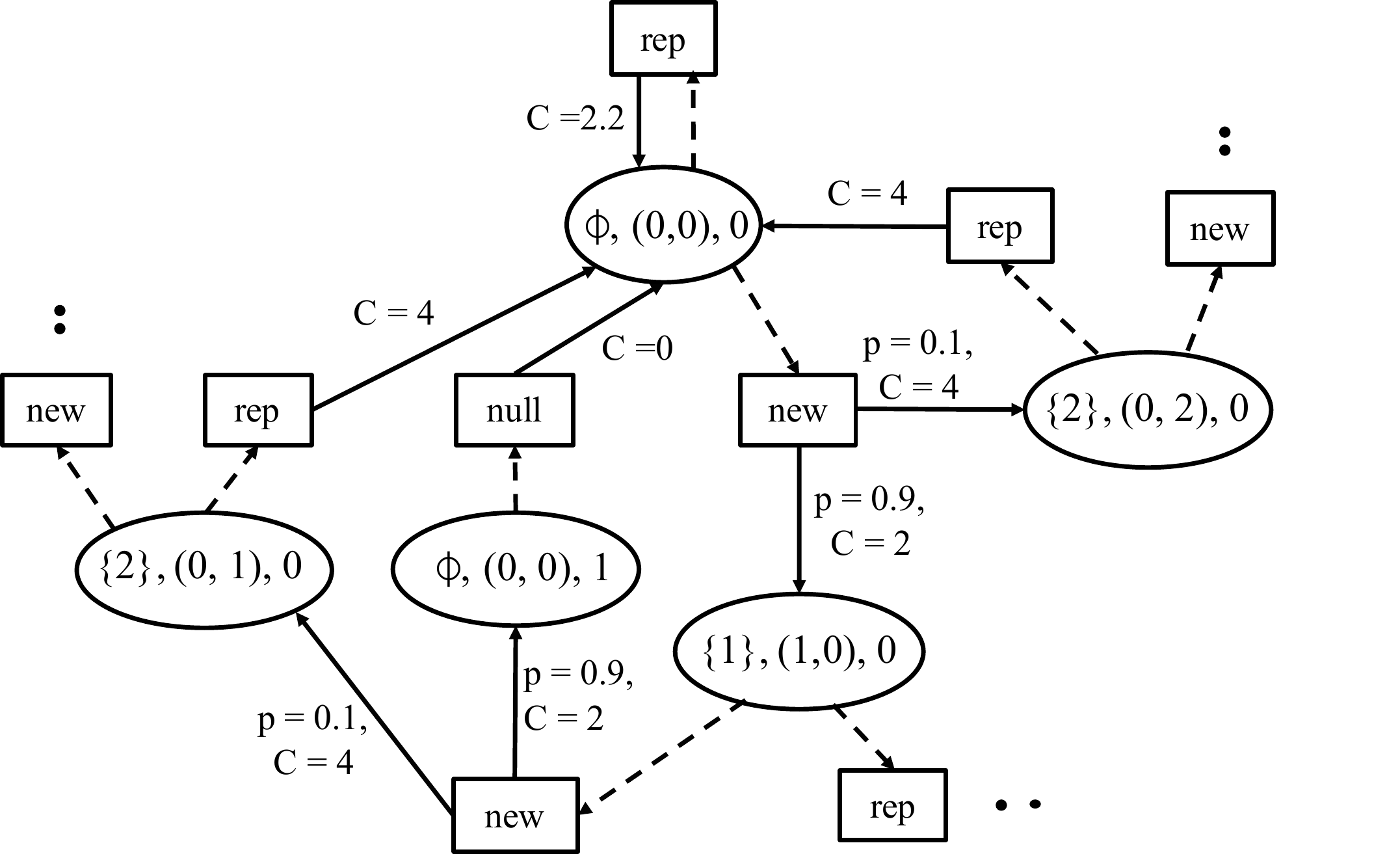}
\caption{Illustration of the MDP for the service distributions in \Cref{subsec:main_ideas}. Dotted arrows correspond to the actions taken from a state and solid arrows lead to the new state resulting from the action. Parts of the MDP resulting from sub-optimal actions are omitted in this figure.\label{fig:mdp_illustration}}
\end{figure}

In states $s = [\mathcal{B}, \mathbf{t}, 0]$, the scheduler can assign new jobs to idle servers (\textbf{\emph{new}}), or replicate existing jobs (\textbf{\emph{rep}}). For example, consider a system of $2$ servers (illustrated in \Cref{fig:mdp_illustration} for the service time distributions in \Cref{exple:adarep_exmple}). In states $[\{2 \}, (0, t), 0]$ or $[\{1\}, (t, 0),0]$ with $t >0$, one server is idle while the other has spent $t$ seconds on its current job. From the state $s = [ \emptyset, (0,0), 0]$ where both servers are idle, the $\textbf{\emph{new}}$ action assigns two new jobs, one to each server, and the $\textbf{\emph{rep}}$ action replicates a new job at both servers. In a system with $K > 2$ servers that is currently in state $s = [\mathcal{B}, \mathbf{t}, 0]$, the action space $\mathcal{A}_s$ contains possible $|\mathcal{B}| + 1$ actions corresponding to replicating one of the $|\mathcal{B}|$ jobs currently running or launching a new job at the idle server.

\subsection{Cost}
The cost $C(s, s', a)$ associated with a transition from state $s$ to $s'$ when action $a$ is taken in state $s$ is defined as the total time spent by the servers in that interval. More formally, $C(s, s', a) = K \E{D_{s \rightarrow s'}}$, where $\E{D_{s \rightarrow s'}}$ is the expected time that elapses between state $s$ to $s'$. For example, consider a system of $K =2$ servers illustrated in \Cref{fig:mdp_illustration} for the service time distributions in \Cref{exple:adarep_exmple}. Starting from $s = [ \emptyset, (0,0), 0]$ where both servers are idle, the $\textbf{\emph{rep}}$ action replicates a new job at both servers. This job takes time $\E{\min(X_1, X_2)} = 2\times 0.1 + 1 \times 0.9 =  1.1$ to depart after which the system transitions back to $s' = [ \emptyset, (0,0), 0]$. The cost associated with this transition is $C(s, s', \textbf{\emph{rep}}) = 2.2$.

The throughput-optimal policy $\SchedPolicy_{n,r}^*$ is the solution to the following cost minimization problem,
\begin{align}
\label{eqn:cost_min_problem}
\SchedPolicy_{n,r}^* =  \arg \min_{\SchedPolicy \in \Pi_{n,r}}  \sum_{j=0}^{\infty} C(s_j, s_{j+1}, \SchedPolicy(s_j)).
\end{align}

As illustrated in \Cref{fig:mdp_illustration}, we observe that this MDP can have a large state-space even for simple service distributions. And more generally, if $X_i$ for any $i$ is a continuous random variable for which the MDP will have a continuous state-space, which becomes even harder to solve. % In general, this MDP is tedious to solve, especially 

%In \Cref{fig:mdp_illustration} we illustrate the MDP for the service distributions in \Cref{exple:adarep_example}. Clearly This MDP can be solved using standard techniques such as value iteration, policy iteration etc. For the system in \Cref{exmp:motiv_example}, the optimal policy ends up being the adaptive replication policy presented in \Cref{exmp:motive_example}. 

%\TODO{Add figures to illustrate the MDP}

%Note that if $X_i$ for any $i$ or $Y$ is a continuous random variable then the possible values taken by $t_i$ is a continuous set, and as a result the MDP will have a continuous state-space.
\section{The MaxRate and AdaRep Replication Policies}
\label{sec:myopic_policies}

As an alternative to solving the MDP, we propose two replication policies, MaxRate and AdaRep. The MaxRate policy, a greedy myopic policy, is defined as follows.

\begin{defn}[MaxRate Policy]
In each state $s = [\mathcal{B}, \mathbf{t}, D_r]$, one server is idle and the replication policies needs to choose action $a \in \mathcal{A}_s$, that is, either launch a new job or replicating one of the jobs currently running at the other servers. The resulting state after taking action $a$ is denoted by $s(a) = [\mathcal{B}(a), \mathbf{t}(a), D_r(a)]$ The MaxRate policy chooses the action $a^*$ that maximizes the instantaneous service rate $\hat{\Thpt}(a)$ which is defined as, %$a \in \mathcal{A}_s$. we evaluate the instantaneous service rate
\begin{align}
\hat{\Thpt}(a) \triangleq \sum_{m=1}^{|\mathcal{B}(a)|} \frac{1}{\E{D_m}} \label{eqn:inst_rate}.
\end{align}
where $|\mathcal{B}(a)|$ is the number of unfinished jobs after taking action $a$, and $\E{D_m}$ is the expected remaining time until the departure of job $m$, assuming that it is not replicated further. %The MaxRate policy chooses action $a^*$ that minimizes $\hat{\Thpt}(a)$.
\end{defn}

The connection between the MaxRate policy and the MDP presented in \Cref{sec:mdp_formu} is as follows. The solution to the MDP $\SchedPolicy_{n,r}^*$ given by \eqref{eqn:cost_min_problem} minimizes the expected computing cost $\E{C}$. As given by \Cref{clm:thpt_in_terms_of_EC} minimizing the expected cost is equivalent to maximizing the throughput $R = \lim_{n \rightarrow \infty} n/T_n$ where $T_n$ is the departure time of the $n^{th}$ job. The MaxRate policy maximizes the instantaneous service rate $\hat{\Thpt}(a)$, a myopic approximation of the throughput $R$ based on the jobs currently present in the system. 
%
% \Cref{defn:computing_cost}. This policy is also the solution of the rate maximization problem
%\begin{align*}
%\SchedPolicy_{n,r}^* =  \arg \max_{\SchedPolicy \in \Pi_{n,r}}  \sum_{j=0}^{\infty} \frac{1}{C(s_j, s_{j+1}, \SchedPolicy(s_j))}.
%\end{align*}
%
% can be equivalently written as 
% minimizes $\E{C(s_j, s_{j+1}, \SchedPolicy(s_j))$, that the instantaneous service rate $\hat{\Thpt}(a)$ of leaving a state $s$ is a proxy for the $1/\E_s{C(s_j, s_{j+1}=s', a)}$, where $C(s_j, s_{j+1}=s', a)$ the cost associated with the transition from state $s$ to $s'$ when action $a$ is taken in state $s$.
%\TODO{Edit this}

\begin{exple}
Consider a two server system, with cancellation delay $\Delta = 0$. Suppose server $1$ becomes idle, and the job assigned to server $2$ has spent time $t_2 > 0$ in service. Let $X_2^{rs} = (X_2 - t_2) | X_2 > t_2$ be the residual computing time. The MaxRate policy launches a replica at server $1$ if
\begin{align}
\frac{1}{\E{\min(X_1, X_2^{rs})}} > \frac{1}{\E{X_1}} + \frac{1}{\E{X_2^{rs}}}. \label{eqn:maxrate_two_servers}
\end{align}
and otherwise it assigns a new job to server $1$.
\end{exple}

\begin{figure}[t]
    \centering
   \includegraphics[width= 0.49\textwidth]{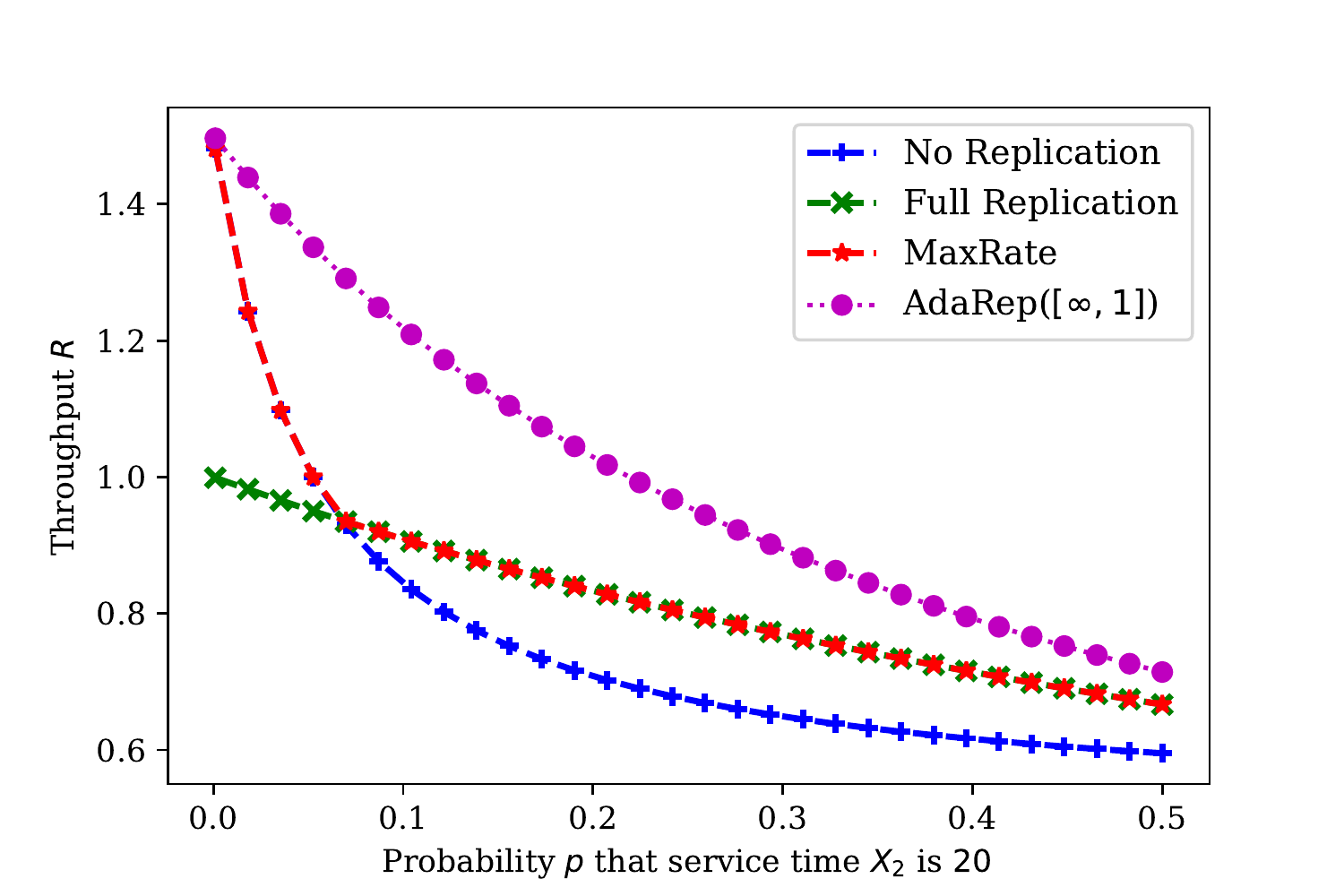}
\caption{For the service distributions in \Cref{exple:adarep_exmple} with different values of $p$, the throughput with the MaxRate policy is a maximum of the throughputs with the FullRep and NoRep policies. The AdaRep policy with a replication threshold of $1$ for jobs originally launched on server $2$ achieves the best throughput. \label{fig:het_comp_max_rate}}
\end{figure}

\begin{figure}[t]
    \centering
   \includegraphics[width= 0.49\textwidth]{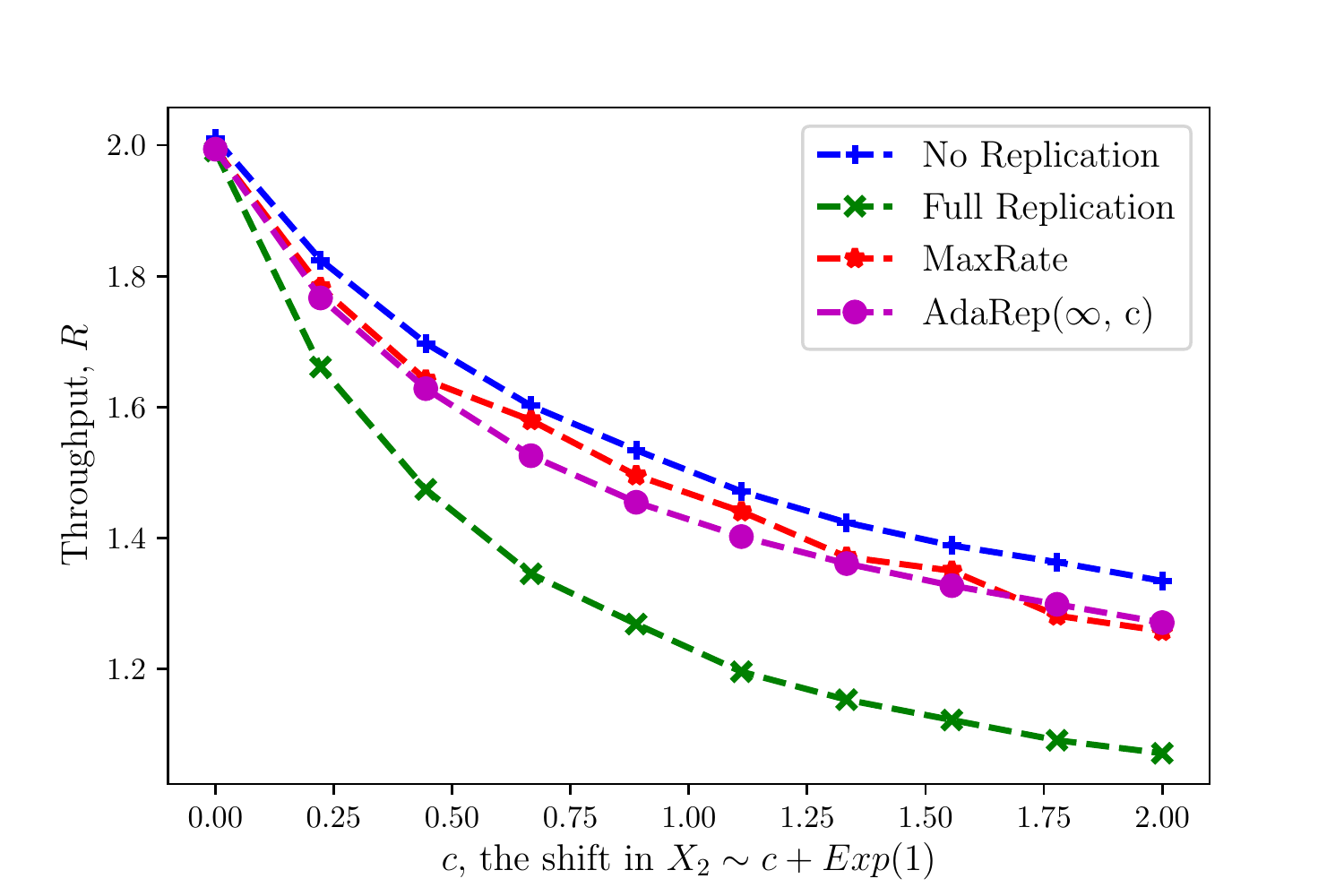}
\caption{For the two-server case with service time distributions $X_1 \sim Exp(1)$ and $X_2 \sim c+\Exp(1)$, we plot the throughputs of MaxRate, AdaRep($[t_{1 \rightarrow 2}, t_{2 \rightarrow 1}]= [\infty, c]$), Full Replication and No replication poilicies versus the initial delay $c$ in the service time $X_2\sim c+\Exp(1)$. The MaxRate and AdaRep policies are close to the no replication policy which yields the best throughput. \label{fig:maxrate_adarep_exp_shifted_exp}}
\end{figure}

The MaxRate policy implicitly finds replication thresholds $t_{i \rightarrow j}$ such that a job running on server $i$ is replicated at server $j$ if it does not finish in $t_{i \rightarrow j}$ seconds. Based on this idea we propose another class of policies called AdaRep($\mathbf{t}$), which is explicitly parametrized by a replication threshold vector $\mathbf{t}$.

\begin{defn}[AdaRep Policy]
Consider a vector of server indices $\mathbf{u} = (j_1, j_2, \dots j_k)$ for $k < K$ such that a job first launched on server $j_1$ was later replicated on $j_2$, $j_3$ and so on. This job is replicated at server $i$ if the job has spent at least $t_{\mathbf{u} \rightarrow i}$ time in service from the time its original copy was launched. Otherwise it assigns a new job to the idle server.
\end{defn}

For example for $K=2$ servers, the vector $\mathbf{t} = \left[ t_{ 1 \rightarrow 2}, \, t_{2 \rightarrow 1} \right]$. In general, choosing the best replication thresholds is a non-trivial problem.  In the next section we propose a method to choose $\mathbf{t}$ for the two-server case.

\begin{figure*}[ht]
\centering
\begin{subfigure}[t]{0.32\textwidth}
    \centering
   \includegraphics[width=1.1\textwidth]{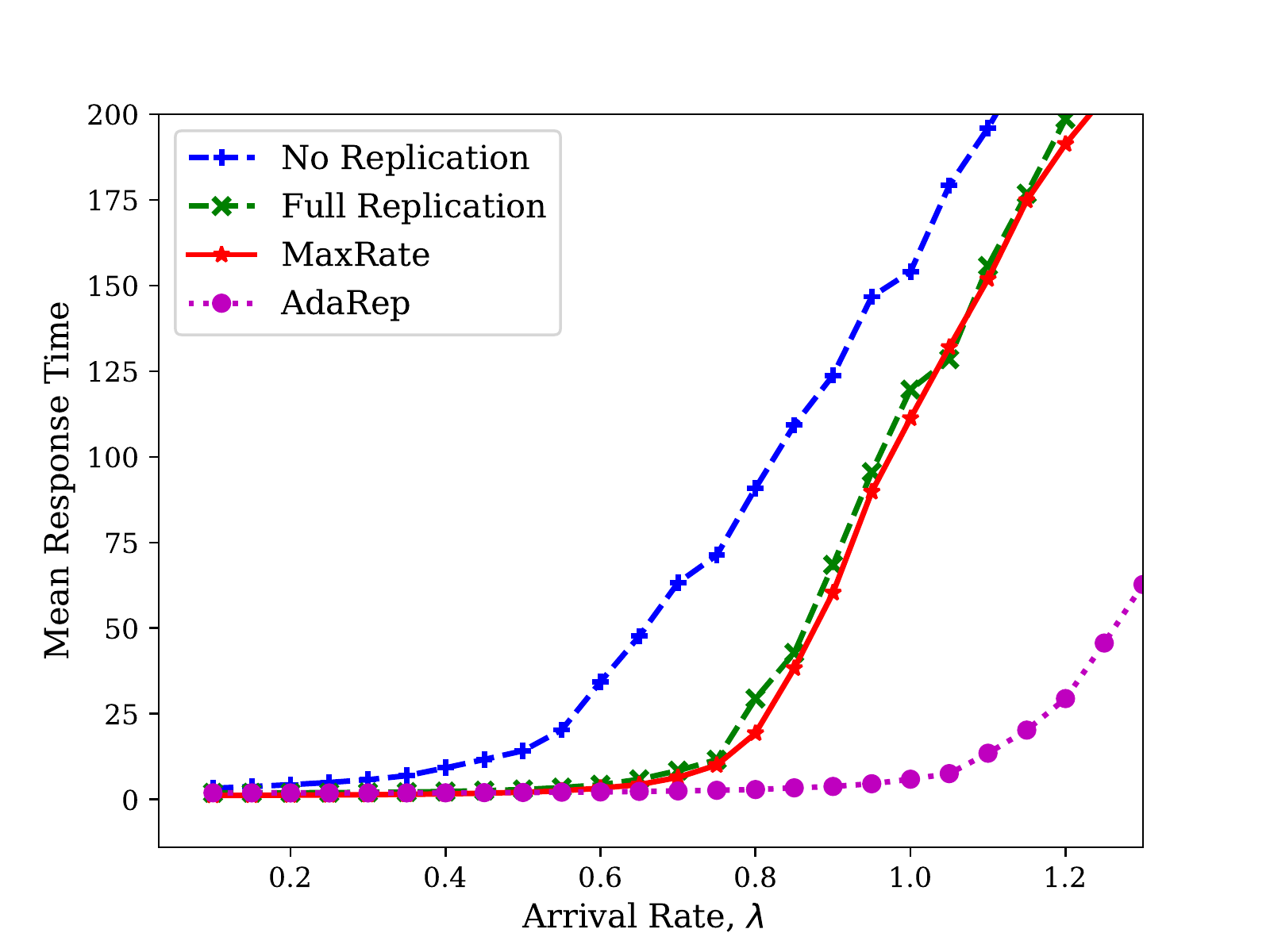}
\caption{$K=2$ servers with service distributions as given in \Cref{exple:adarep_exmple} with $p = 0.1$. The replication thresholds used in the AdaRep policy are $(t_{1 \rightarrow 2}, t_{2 \rightarrow 1}) = (\infty, 1)$.
\label{fig:example_2_server}}
\end{subfigure}
~
\begin{subfigure}[t]{0.32\textwidth}
    \centering
   \includegraphics[width=1.1\textwidth]{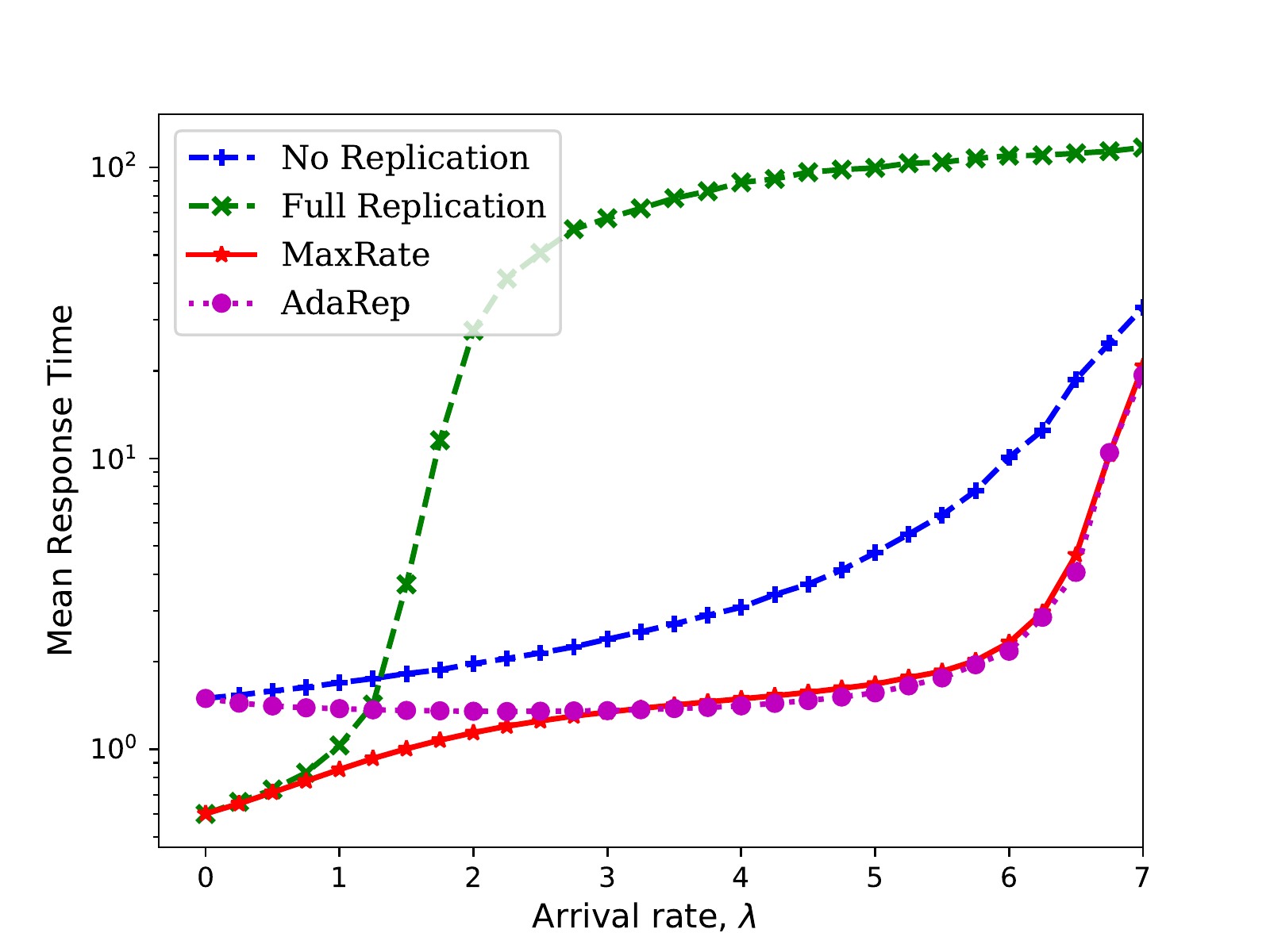}
\caption{$K=10$ homogeneous servers with shifted exponential service times $X \sim 0.5 + Exp(1)$. The AdaRep replication thresholds for launching additional replicas of a job are $(0.1, 0.2, \dots, 0.9)$, that is, we launch the $i^{th}$ additional replica if the original copy of the job spends at least $0.1 \times i$ units of time in service.  \label{fig:shifted_exp_10_server}}
\end{subfigure}
~
\begin{subfigure}[t]{0.32\textwidth}
    \centering
   \includegraphics[width=1.1\textwidth]{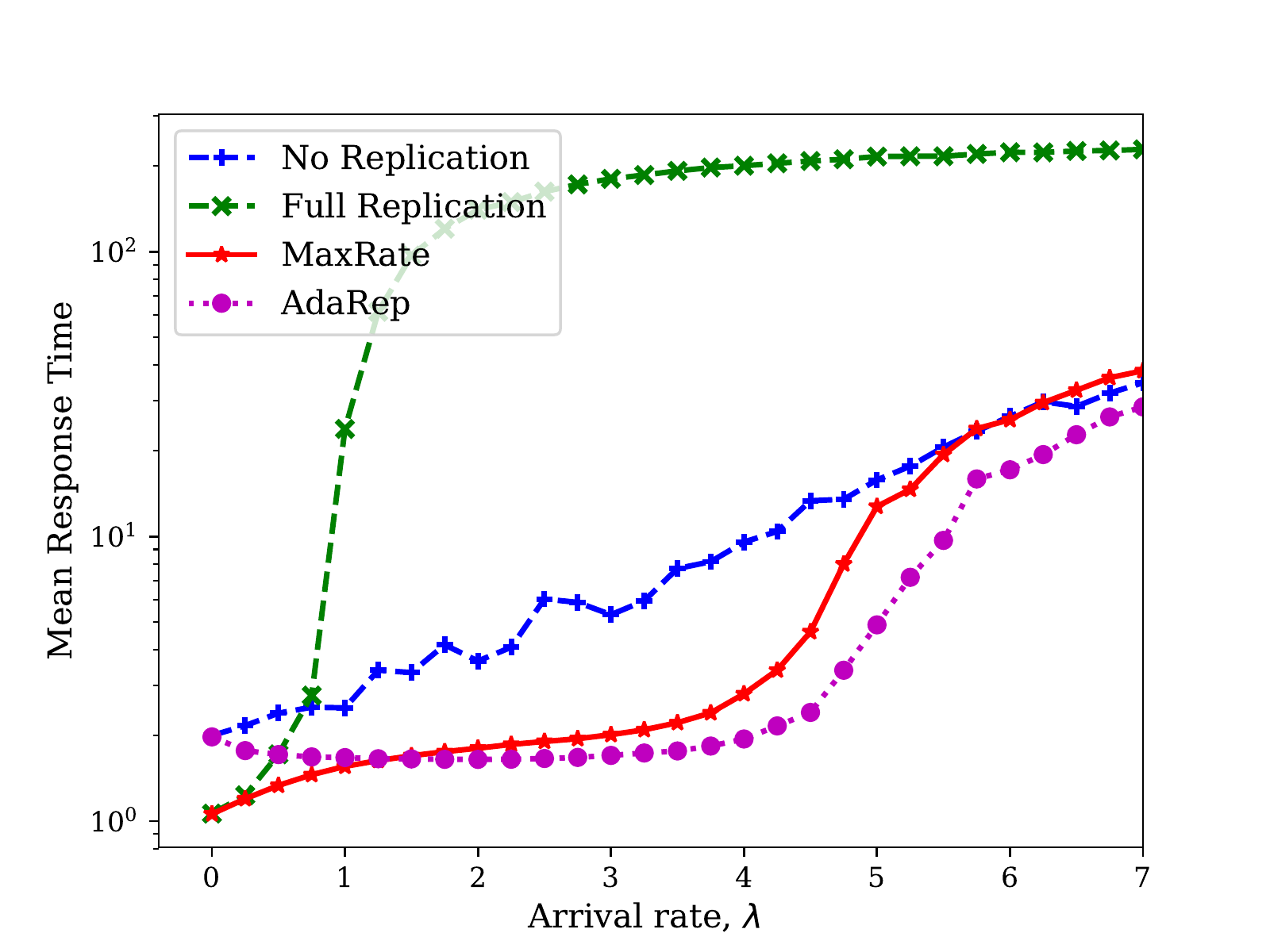}
\caption{$K=10$ homogeneous servers with Pareto $X \sim Pareto(1,2)$. The AdaRep replication thresholds for launching additional replicas of a job are $(0.1, 0.2, \dots, 0.9)$, that is, we launch the $i^{th}$ additional replica if the original copy of the job spends at least $0.1 \times i$ units of time in service.  \label{fig:pareto_10_server}}
\end{subfigure}
\caption{Mean response times versus arrival rate $\lambda$ for the No Replication, Full Replication, MaxRate and AdaRep policies. \label{fig:response_time_plots}}
\end{figure*}

%\subsection{Throughput Comparison and Scaling}
\Cref{fig:het_comp_max_rate} illustrates the MaxRate and AdaRep policies in comparison with the FullRep and NoRep policies for the service distributions in \Cref{exple:adarep_exmple}. Observe that the throughput of the MaxRate policy is the maximum of the throughputs of the NoRep and FullRep policies. The AdaRep($[\infty, 1]$) policy proposed in \Cref{exple:adarep_exmple} gives a throughput $R = 1.2185$ when $p = 0.1$, which is significantly higher than $R = 0.909$ achieved by the greedy myopic MaxRate policy. However, unlike MaxRate where the replication thresholds are implicitly found by maximizing the instantaneous job departure rate, the throughput of the AdaRep policy is highly sensitive to the choice of the replication thresholds. %. Thus, MaxRate is not close to optimal, which is not surprising because it is greedy, and oblivious to the system state resulting the action.  

\Cref{fig:maxrate_adarep_exp_shifted_exp} illustrates the MaxRate and AdaRep policies in comparison with the FullRep and NoRep policies for two servers with exponential $X_1 \sim Exp(1)$ and shifted exponential $X_2 \sim c+Exp(1)$ service time distributions (for a constant $c \geq 0$) respectively. Due to the memoryless property of the exponential distribution, when $c = 0$, the no replication and full replication policies give the same throughput. When $c > 0$, full replication gives strictly lower throughput than no replication. Observe that the MaxRate policies tries to dynamically emulate NoRep. AdaRep with a replication threshold $t_{2 \rightarrow 1} = c$ for jobs originally launched in server $2$ gives lower throughput than NoRep because the optimal replication thresholds are $[t_{1 \rightarrow 2}, t_{2 \rightarrow 1}] = [\infty, \infty]$ in this case.

%
%\TODO{A plot of throughput versus $\Delta$ comparing the no replication, full replication, maxrate and Adarep policies for the two server example where both servers have shifted exponential service time distributions}

\subsection{Mean Response Time in Low and Moderate Traffic Regimes}
\label{sec:response_time}

Although the focus of this paper is to find throughput-optimal replication policies, the MaxRate and AdaRep policies proposed above work very well in the low and moderate traffic regimes as we show via simulations below. The simulation setting is as follows. We consider Poisson job arrivals with rate $\lambda$ into the central queue shown in \Cref{fig:het_servers_sys_model} instead of assuming that the queue is saturated with jobs. Unlike the saturated central queue case considered so far, where a server is assigned a new job (or a replica of an existing job) as soon as it becomes idle, servers now remain idle when there are no jobs in the central queue. We then record the mean response time  (waiting time in queue plus service time) experienced by jobs averaged over $100$ simulation runs with $1000$ jobs each.

\Cref{fig:example_2_server} shows a comparison of the mean response time experienced by jobs for the No Replication, Full Replication, MaxRate and AdaRep($[\infty, 1]$) policies for the two server case with service distributions as given in \Cref{exple:adarep_exmple} with $p = 0.1$. The cancellation delay is assumed to be $\Delta = 0$. In addition to boosting the throughput in high-traffic, MaxRate and AdaRep also result in faster response times in the low traffic regime. As $\lambda$ increases, observe that MaxRate transitions from full replication to no replication. The AdaRep policy gives the lowest response time in all traffic regimes.

%\begin{figure}[t]
%    \centering
%   \includegraphics[width= 0.49\textwidth]{spe_2server.pdf}
%\caption{Mean response times versus arrival rate $\lambda$ for the No Replication, Full Replication, MaxRate and AdaRep($[\infty, 1]$) policies for the two server case with service distributions as given in \Cref{exple:adarep_exmple} with $p = 0.1$. 
%\label{fig:response_time_2_server}}
%\end{figure}
%
%\begin{figure}[t]
%    \centering
%   \includegraphics[width=0.49\textwidth]{sh-exp_10server.pdf}
%\caption{Mean response times versus arrival rate $\lambda$ for the No Replication, Full Replication, MaxRate and AdaRep policies. The system has $K=10$ homogeneous servers with shifted exponential service times $X \sim 0.5 + Exp(1)$. \label{fig:response_time_10_server}}
%\end{figure}

\Cref{fig:shifted_exp_10_server}  shows a comparison of the mean response time experienced by jobs for the No Replication, Full Replication, MaxRate and AdaRep policies for $K=10$ homogeneous servers with shifted exponential service times $X \sim 0.5 + Exp(1)$. The AdaRep replication thresholds for launching additional replicas of a job are $(0.1, 0.2, \dots, 0.9)$, that is, we launch the $i^{th}$ additional replica if the original copy of the job spends at least $0.1 \times i$ units of time in service. The cancellation delay is assumed to be $\Delta = 0$. In addition to boosting the throughput in high-traffic, MaxRate and AdaRep also result in faster response times in the low traffic regime. In very low traffic, MaxRate launches replicas at all the idle servers in order to greedily maximize the instantaneous job departure rate, and thus, its throughput resembles that of the full replication policy. As $\lambda$ increases, both MaxRate and AdaRep replicate fewer times and come closer to no replication, which is throughput-optimal in heavy traffic. A similar trend is observed in \Cref{fig:pareto_10_server} for $K=10$ homogeneous servers with Pareto service times $X \sim Pareto(1,2)$. Here, the AdaRep policy performs better than MaxRate in the moderate traffic regime.

%\begin{figure}[t]
%    \centering
%   \includegraphics[width= 0.49\textwidth]{pareto_10server.pdf}
%\caption{For the service distributions in \Cref{exple:adarep_exmple} with different values of $p$, the throughput with the MaxRate policy is a maximum of the throughputs with the FullRep and NoRep policies. \label{fig:het_comp_max_rate}}
%\end{figure}

%\TODO{A simulation plot of $\E{T}$ versus $\lambda$ comparing the no replication, full replication, maxrate and Adarep policies for the two server example}

%\TODO{A simulation plot of $\E{T}$ versus $\lambda$ comparing the no replication, full replication, maxrate and Adarep policies for 10 homogeneous servers with shifted exponential service times $X \sim 0.5 + Exp(1)$.} 

%\TODO{A simulation plot of $\E{T}$ versus $\lambda$ comparing the no replication, full replication, maxrate and Adarep policies for 10 homogeneous servers with Pareto service time}

%\TODO{Prove that there is no loss of generality in restricting our attention to Adarep policies -- Cannot prove this, it is not true}

%\TODO{Mention that although these policies seem to require online computation, the computation can actually be done offline}
%
\section{Upper Bounds on the Service Capacity}
\label{sec:converse_bnd}
In order to quantify the optimality gap of the MaxRate and AdaRep policies proposed above and understand the limits of the service capacity of a multi-server system with job replication, we now provide two fundamental throughput upper bounds. The first bound is for a system of $K=2$ heterogeneous servers, and the second is for a system of $K > 2$ homogeneous servers. The derivations of these bounds use two different techniques to construct a genie system whose throughput is always better than the system under consideration. 
%
%To quantify the optimality gap of a policy without solving the MDP, we need an upper bound on $R_{n,r}^*$. In this section we propose such a converse bound on $R_{n,r}^*$. Drawing insights from this bound, we also propose a method to choose the replication thresholds of the adaptive replication policy.

\subsection{Upper Bound for Two Heterogeneous Servers}
Recall that in our problem formulation, jobs can be replicated only at time instants when one or more servers become idle. To find the upper bound on $R_{n,r}^*$, we consider a system where the scheduler is also allowed to pause ongoing jobs.

\begin{defn}[The Pause-and-Replicate System]
A job can be replicated at any server where it is not already running by pausing the ongoing job on that server. The paused job is resumed when the replica is either served or canceled. 
\end{defn}

For the example shown in \Cref{fig:adarep_exmple}, the pause-and-replicate system can pause job $g$ at time $7$ to run a replica of job $h$, and resume job $g$ afterwards. Both $g$ and $h$ will then finish at time $9$, which is $1$ second faster than with the AdaRep policy without job pausing.
\begin{clm}
\label{clm:converse_bnd}
The service capacity $\Thpt_{p,r}^*$ of the pause-and-replicate system is an upper bound on the service capacity $\Thpt_{n,r}^*$ of the original system.
\end{clm}
\begin{proof}
The set of feasible policies $\Pi_{n,r}$ is a subset of $\Pi_{p,r}$, the set of policies in the pause-and-replicate framework. Thus,
\begin{align}
\Thpt_{p,r}^* &= \max_{\pi \in \Pi_{p,r}} \Thpt (\pi) \geq \max_{\pi \in \Pi_{n,r}} \Thpt (\pi) = \Thpt_{n,r}^*. \label{eqn:genie_thpt_3}
\end{align}
\end{proof}

In the pause-and-replicate framework, the AdaRep($\mathbf{t}$) policy can replicate a job exactly after $t_{\mathbf{u} \rightarrow i}$, instead of waiting for server $i$ to become idle. In \Cref{thm:two_server_thpt_bnd} below, we obtain a closed-form expression for the throughput $\Thpt_{p,r}(\mathbf{t})$ of the AdaRep policy for $K=2$ servers. We show that there is no loss of generality in focusing on AdaRep policies. Thus, in the two-server case, the converse bound $R_{p,r}^* = R_{p,r}(\mathbf{t^*})$, the throughput of the best AdaRep policy.

\begin{figure}[t]
    \centering
   \includegraphics[width= 0.49\textwidth]{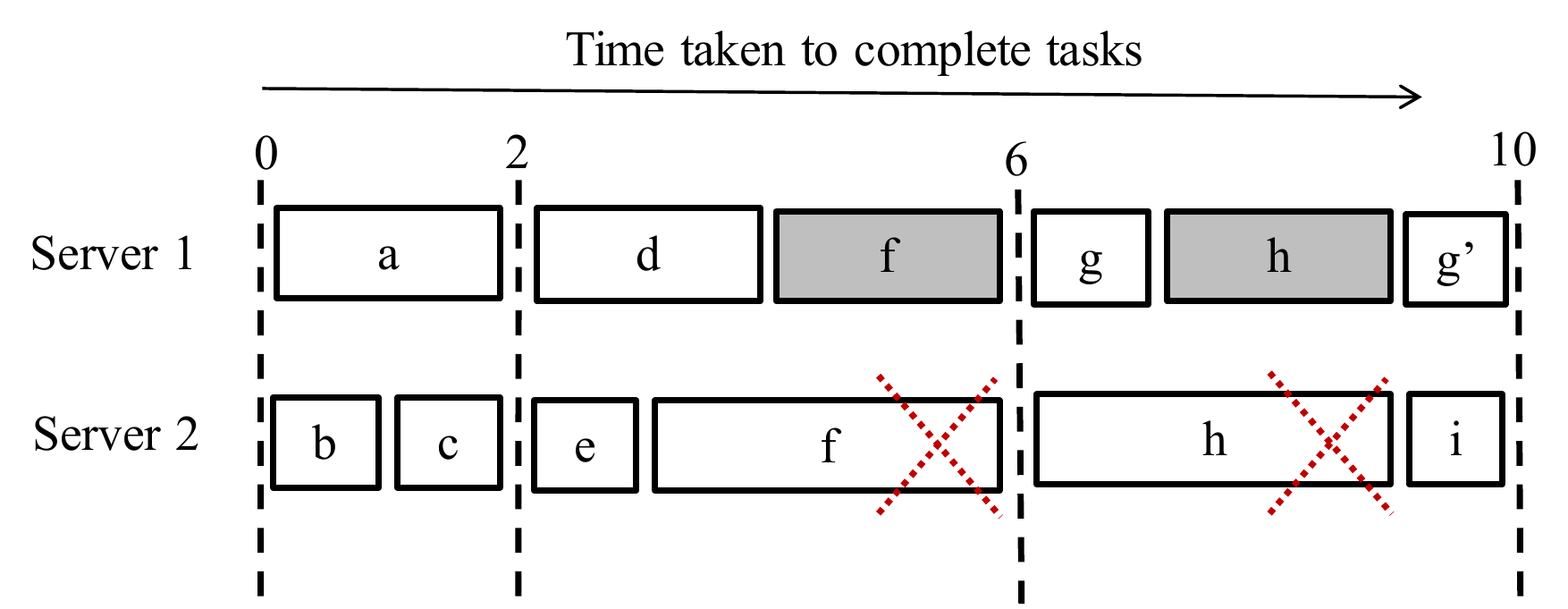}
\caption{Illustration of the optimal replication policy in the pause-and-replicate framework. The service distributions are as given in \Cref{exple:adarep_exmple}. Due to the ability to pause jobs this policy completes more jobs than the AdaRep policy in \Cref{fig:adarep_exmple}. \label{fig:pause_and_rep_exmple}}
\end{figure} 

\begin{thm}
\label{thm:two_server_thpt_bnd}
The throughput $\Thpt_{p,r}( \mathbf{t})$ of AdaRep($\mathbf{t}= \left[ t_{ 1 \rightarrow 2}, \, t_{2 \rightarrow 1} \right]$) in the pause-and-replicate framework can be expressed as follows. For $t_{ 1 \rightarrow 2} > 0$ and $t_{ 2 \rightarrow 1} > 0$,
\begin{align}
\Thpt_{p,r}(\mathbf{t})
&= \frac{\E{\ServTime_1^{tr}(t_{1 \rightarrow 2})} + \E{\ServTime_2^{tr}(t_{2 \rightarrow 1})}}{\E{\ServTime_1^{tr}(t_{1 \rightarrow 2})} \E{\ServTime_2^{tr}(t_{2 \rightarrow 1})} (1 + \gamma_{1 \rightarrow 2} +  \gamma_{2 \rightarrow 1}) }\label{eqn:thpt_bnd_two_server}
\end{align}
where,
\begin{align}
\gamma_{t_{1 \rightarrow 2}} &\triangleq \frac{\Pr(X_1 > t_{1 \rightarrow 2}) (\Delta + \E{\min(X_1^{rs}(t_{1 \rightarrow 2}), X_2)})}{\E{\ServTime_1^{tr}(t_{1 \rightarrow 2})}} \\
\gamma_{ t_{2 \rightarrow 1}} &\triangleq \frac{\Pr(X_2 > t_{2 \rightarrow 1}) (\Delta + \E{\min(X_1, X_2^{rs}(t_{2 \rightarrow 1})})}{\E{\ServTime_2^{tr}(t_{2 \rightarrow 1})}},
\end{align}
and $\ServTime_i^{tr}(\tau)= \min(\ServTime_i, \tau)$, the truncated part of $X_i$, and $\ServTime_i^{rs}(\tau) = (\ServTime_i-\tau) | (\ServTime_i > \tau)$, the residual service time after $\tau$ seconds of service.

If $t_{ 1 \rightarrow 2} = 0$ or $t_{ 2 \rightarrow 1} = 0$,
\begin{align}
\Thpt_{p,r}( \mathbf{t}) &= \frac{1}{\Delta + \E{\min(X_1, X_2)}} \label{eqn:thpt_bnd_two_server_full_rep}.
\end{align}
\end{thm}

The proof is given in the Appendix. In \Cref{coro:two_server_thpt_bnd} below we give the throughput expression for the special case where $t_{ 1 \rightarrow 2}$ set to infinity.

\begin{coro}
\label{coro:two_server_thpt_bnd}
The throughput $\Thpt_{p,r}(\mathbf{t} = \left[ \infty, \, t_{2 \rightarrow 1} \right])$ of the two-server pause-and-replicate system is
\begin{align}
\Thpt_{p,r}(t_{2 \rightarrow 1}) &= \frac{\E{X_2^{tr}(t_{2\rightarrow 1})}}{\E{X_2^{ac}}} \left(\frac{1}{\E{X_1}}\right) + \frac{1}{\E{X_2^{ac}}} 
 \label{eqn:thpt_bnd_two_server}
\end{align}
where, $\E{X_2^{tr}} = \min(X_2, t_{2\rightarrow 1})$, is the truncated part of $X_2$, and $\E{X_2^{ac}}$ is the effective service time of jobs launched on server $2$.
\begin{align}
\E{X_2^{ac}} = \E{X_2^{tr}}  + \Pr(X_2 > t_{2 \rightarrow 1}) (\Delta + \E{\min(X_1, X_2^{rs}}),
\end{align} 
where $\ServTime_2^{rs}(t_{2\rightarrow 1}) = (\ServTime_2- t_{2 \rightarrow 1} | (\ServTime_2 > t_{2 \rightarrow 1}))$, the residual service time after time $t_{2 \rightarrow 1}$ of service.
\end{coro}

Here is an intuitive explanation of the throughput in \eqref{eqn:thpt_bnd_two_server}. Since server $2$ is never paused, its throughput of server $2$ is $1/\E{X_2^{ac}}$, where $\E{X_2^{ac}}$ accounts for the reduction in service time due to replication of jobs. For server $1$, the throughput is $\zeta/\E{X_1}$, where $\zeta = \E{X_2^{tr}}/\E{X_2^{ac}}$, the fraction of time server $1$ is not paused.

To maximize the throughut we can find $t_{2 \rightarrow 1}$ that maximizes \eqref{eqn:thpt_bnd_two_server}. For example, for the service distributions in \Cref{exple:adarep_exmple}, $t^*_{2 \rightarrow 1} = 1$. Thus, if a job does not finish in $1$ seconds on server $2$, we launch a replica on server $1$ by pausing its ongoing job. This policy is illustrated in \Cref{fig:pause_and_rep_exmple}.

\begin{figure}[t]
    \centering
   \includegraphics[width= 0.49\textwidth]{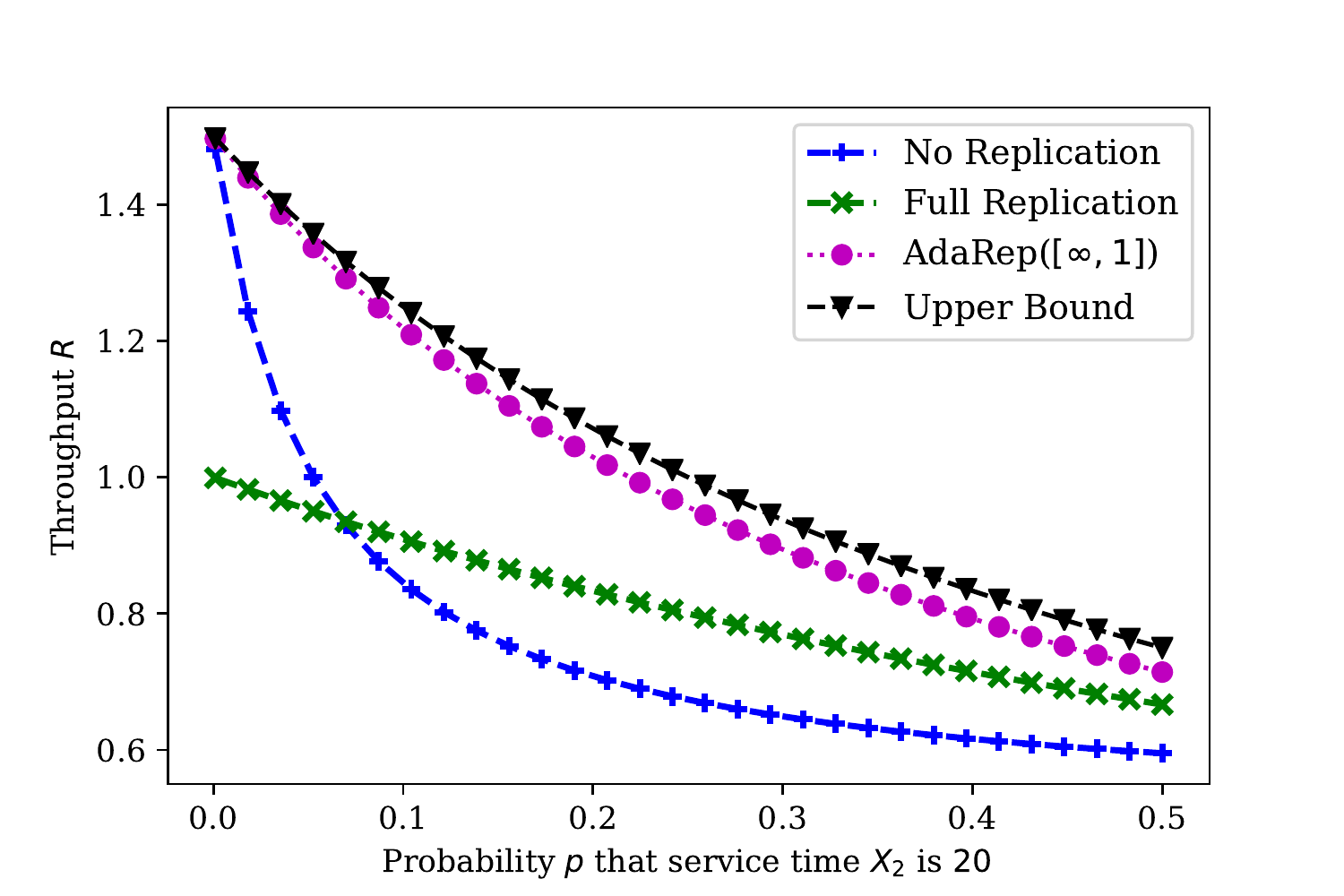}
\caption{Illustration of the upper bound on the service capacity $\Thpt^*_{n,r}$, along with the NoRep, FullRep and AdaRep policies. The service distributions are as defined in \Cref{exple:adarep_exmple}, with $p$ varying along the x-axis. The AdaRep policy with carefully chosen replication thresholds gives throughput that is close to the upper bound. \label{fig:het_comp_adarep_bnd}}
\end{figure}

\begin{lem}
\label{lem:single_pt_optimality}
For $K=2$ servers, there is no loss of generality in focusing on AdaRep policies to find the optimal throughput $R_{p,r}^*$ in the pause-and-replicate framework. That is, $R^*_{p,r} =  \max_{\mathbf{t}} R_{p,r}(\mathbf{t})$.
\end{lem}

The proof is given in the appendix. Based on this throughput upper bound, we propose an adaptive replication policy $\pi_{AdaRep}$. This policy tries to emulate the best AdaRep thresholds $\mathbf{t}^*$, under the limitation that it cannot pause ongoing jobs to launch replicas. Recall that we already saw this policy in \Cref{exple:adarep_exmple}. Now we define it more generally for any service distributions. In \Cref{fig:het_comp_adarep_bnd} we plot the throughput achieved by the AdaRep policy, alongwise the upper bound given by maximizing \eqref{eqn:thpt_bnd_two_server} over $t_{2 \rightarrow 1}$. The service distributions are same as \Cref{exple:adarep_exmple} with $p$ varying along the x-axis. We observe that the AdaRep policy comes closest to the upper bound, with the gap resulting from its inability to pause jobs.

Generalizing this pause-and-replicate system based bounding technique to more than two servers is a difficult and non-trivial problem. This is because there can be deadlock situations where a job being run on a server $a$ cannot be replicated at server $b$ because server $c$ had already paused server $b$ to replicate its own current job. This makes it difficult to find a closed-form expression of the throughput as we did in the two server case in \Cref{thm:two_server_thpt_bnd}. Below, we present a different upper bounding technique for the case of $K$ homogeneous servers.

\subsection{Upper Bound for $K$ homogeneous servers}
Going beyond the case of two heterogeneous servers, we now present an upper bound on the service capacity for a system of $K$ homogeneous servers. This bound holds for any service time distribution $X \sim F_X$ and cancellation delay $\Delta$. %It can be generalized to $K$ heterogeneous servers by taking all permutations of relative task start times, but for brevity purposes we only present the homogeneous server case here. 
%Note that the technique used to find this upper bound is conceptually different from the pause-and-replicate system presented for the two-server case presented above. We expect the pause-and-replicate system method to yield a tighter bound in the heterogeneous two server case. 

\begin{thm}[Throughput Upper Bound for $K$ Homogeneous Servers]
\label{thm:general_upper_bnd}
For a system of $K$ homogeneous servers with service times $X  \sim F_X$ that are i.i.d.\ across jobs and servers, and cancellation delay $\Delta$, the service capacity $R^{*}_{n,r}$ is bounded as follows
\begin{align}
R^{*}_{n,r} &\leq  \frac{K}{ \min_{0 \leq t_2 \leq \dots \leq t_{K}}  \mathbb{E}_X[C(t_2, \dots, t_K)]} 
 \label{eqn:general_upper_bnd}
\end{align}
where 
\begin{align}
 \mathbb{E}_X[C (t_2, \dots, t_K)] &=  \sum_{k=1}^{K}  \mathbb{E}_X[ (S - t_k)^+ ] + \nonumber \\
& \hspace{0.3 cm} \Delta \left(\sum_{k=2}^{K} \E{\mathbbm{1}(t_k < S)} + \E{\mathbbm{1}(t_2 < S)} \right) \label{eqn:C_in_terms_of_t_k}
\end{align}
where $t_2 \leq t_3, \leq \dots t_K$ are the start times of the replicas of a job relative to the first copy which starts at time $t_1 = 0$, and $(x)^+ = \max(0,x)$. The service time $S$ of a job is the time from the start of the earliest copy until any one of the replicas is served, that is,
\[ S = \min(X^{(1)}, X^{(2)}+ t_2, \dots,  X^{(K)}+t_K), \]
where $X^{(k)} \sim F_X$.
\end{thm}

The proof is given in the Appendix. Now we look at some special cases to give an intuitive understanding  of this upper bound. When the service time of each server is deterministic, that is, $X = c$, observe that the throughput upper bound is $K/c$, which is achieved by the no replication policy. On the other hand, if we have two homogeneous servers with exponential service times $X^{(1)}, X^{(2)} \sim Exp(\mu)$ and cancellation delay $\Delta > 0$,  then we have
\begin{align}
S &= X^{(1)} \mathbbm{1}_{X^{(1)} \leq t_2} + (t_2 + \Exp(2 \mu)) \mathbbm{1}_{X^{(1)}>t_2} \\
C(t_2) &= X^{(1)} \mathbbm{1}_{X^{(1)} \leq t_2} +(t_2 + 2\Exp(2 \mu)+2\Delta) \mathbbm{1}_{X^{(1)}>t_2}\\
\E{C(t_2)} &= \frac{1}{\mu} + 2 \Delta e^{-t_2 \mu}
\end{align}
To minimize $\E{C(t_2)}$ we need to set $t_2$ to $\infty$, and thus we get $R^{*}_{n,r} \leq 2 \mu$. This implies that the no replication policy is the best for exponentially distributed service times when there is a non-zero cancellation delay.

\section{Concluding Remarks}
\label{sec:conclu}
The traditional view of a multi-server system is that its service capacity or maximum possible throughput is equal to the sum of the service rates. However, when we employ job replication to overcome variability in service time, there is a paradigm shift in this notion of service capacity -- redundancy, when used effectively, can lead to a synergistic combination of servers such that the overall throughput is greater than the sum of the service rates of individual servers. Motivated by the idea that job replication can boost the throughput of multi-server system, this paper aims to find the maximum possible throughput. We first tackle the simpler case of upfront replication and determine the optimal number of replicas that maximize the throughput. Next, we consider gradual launch of additional replicas and propose two replication policies: MaxRate, a myopic policy based on an MDP formulation of the problem, and AdaRep, a tunable threshold-based replication policy. These policies are effective even in the low and moderate traffic regimes as demonstrated by simulation results presented in this paper. We also obtain upper bounds on the service capacity to understand the fundamental limits of the achievable throughput. 

The main contribution of this paper is to demonstrate how replication can not only cope with service variability, but also make more efficient use of computing resources. Generalizations of the system model include allowing job killing and accounting for data locality constraints. Another future direction is analyze the mean response time of delayed job replication policies and finding the optimal policy in the low and moderate traffic regime. Designing online learning-based replication policies that can function in a system where the service time distributions are unknown or time-varying is also an interesting open problem.

\section*{Acknowledgments}
The authors thank Emina Soljanin, Devavrat Shah, Gregory Wornell, Isaac Grosof, Mor Harchol-Balter and Samarth Gupta for fruitful discussions. We are also extremely grateful to anonymous reviewers for their insightful comments that helped improve this paper.
\appendix
\label{app:appendix}

\begin{proof}[Proof of \Cref{clm:work_conserving_optimality}]
Consider a non-work-conserving scheduling policy $\SchedPolicy_{nwc}$ which results in job departure times $T_1(\SchedPolicy_{nwc}) \leq \dots \leq T_n(\SchedPolicy_{nwc})$. Construct a work-conserving $\SchedPolicy_{wc}$ that follows all the actions of $\SchedPolicy_{nwc}$, except the idling of servers. For example, consider a set of $r \geq 1$ servers that become idle at times $h_1, h_2, \dots , h_r$ respectively. If $\SchedPolicy_{nwc}$ launches replicas of a job $i$ on these servers at times $h_1 + \epsilon_1, h_2 + \epsilon_2, \dots, h_r +\epsilon_r$, where $\epsilon_j \geq 0$ are the idle times, then $\SchedPolicy_{wc}$ starts the replicas at times $h_1, h_2, \dots , h_r$ instead.

We use induction to prove that $T_i (\SchedPolicy_{nwc}) \geq T_i (\SchedPolicy_{wc})$ for all $1 \leq i \leq n$. In both policies, all servers are available for job assignment at time $0$. The departure time of the first job is 
\begin{align}
T_1 (\SchedPolicy_{nwc}) &=  \min(X_1 + \epsilon_1, X_2 + \epsilon_2, \dots X_r + \epsilon_r) \\
&\geq \min(X_1, X_2, \dots X_r)\\
& = T_1 (\SchedPolicy_{wc}).
\end{align}
This is the induction base case. For the induction hypothesis, assume that for all $i \leq n-1$, $T_i (\SchedPolicy_{nwc}) \geq T_i (\SchedPolicy_{wc})$. We now prove that $T_{n}(\SchedPolicy_{nwc}) \geq T_{n}(\SchedPolicy_{wc})$.
Suppose $\SchedPolicy_{nwc}$ assigns job $n$ to $r \geq 1$ servers. The times $h_1, h_2, \dots, h_r$ when these servers become idle belong to the set $\{0, T_1(\SchedPolicy_{nwc}), \dots T_{n-1}(\SchedPolicy_{nwc})\}$, the departure times of previous jobs. By the induction hypothesis, with $\SchedPolicy_{wc}$ the servers become idle earlier at times $g_1, g_2, \dots, g_r$ where $g_j \leq h_j$ for all $1 \leq j \leq r$. Thus,
\begin{align}
T_{n} (\SchedPolicy_{nwc}) &= \min(X_1 +h_1 + \epsilon_1, X_2 + h_2 +\epsilon_2,  \dots , \nonumber\\
&\hspace{1.75cm} X_r + h_r +\epsilon_r) \\
&\geq  \min(X_1 +h_1 ,  X_2 + h_2, \dots X_r + h_r)\\
&\geq  \min(X_1 +g_1, X_2 + g_2, \dots X_r + g_r) \\
&= T_{i+1}(\SchedPolicy_{wc})  
\end{align}

Thus, by induction, $T_n (\SchedPolicy_{nwc}) \geq T_n (\SchedPolicy_{wc})$ for any $n \in \mathcal{N}$. Hence by \eqref{eqn:thpt_in_terms_of_total_delay}, $\Thpt (\SchedPolicy_{nwc}) < \Thpt (\SchedPolicy_{wc})$.
\end{proof}

\begin{proof}[Proof of \Cref{thm:upfront_rep}]
Incoming jobs are replicated at any one super-server, and the replicas are canceled as soon as one copy is served. Thus, the total time spent by each server in super-server $\mathcal{S}_j$ on a job is $\min_{l \in \mathcal{S}_j} X_l + \Delta$. The throughput of that super-server is
\begin{align}
R_{\mathcal{S}_j} &= \frac{1}{\E{\min_{l \in \mathcal{S}_j} X_l } + \Delta}.
\label{eqn:thpt_upfront_rep}
\end{align}
The overall throughput is the sum of the throughputs of the super-servers $\mathcal{S}_1, \mathcal{S}_2, \dots, \mathcal{S}_h$, and is given by \eqref{eqn:thpt_upfront_rep}.
\end{proof}

\begin{proof}[Proof of \Cref{thm:scaling_upfront_K_servers}]
Let the number of servers in server $i's$ group be denoted by $r_i$. For example if $K=5$ are divided into two groups of $3$ and $2$, then $r_1 = r_2 = r_3 = 3$ and $r_4 = r_5 = 2$. The throughput of a group with $r_i$ servers is $1/(\E{X_{1:r_i}} + \Delta)$. If we normalize by the number of servers, the throughput per server is $1/r_i (\E{X_{1:r_i}} + \Delta)$. Summing this over all servers we have,
\begin{align}
R_{UpFr} &= \sum_{i=1}^{K} \frac{1}{r_i (\E{X_{1:r_i}} + \Delta)} \\
&\leq  \sum_{i=1}^{K}  \frac{1}{r^* (\E{X_{1:r^*}} + \Delta)} \label{eqn:opt_thpt_upfront_rep_2} \\
&=  \frac{K}{r^* (\E{X_{1:r^*}} + \Delta)}
\end{align} 
If $r^*$ divides $K$, then dividing servers into groups of $r^*$ servers each gives equality in \eqref{eqn:opt_thpt_upfront_rep_2} above.
\end{proof}

\begin{figure}[t]
\centering
 \includegraphics[width= 3.5 in]{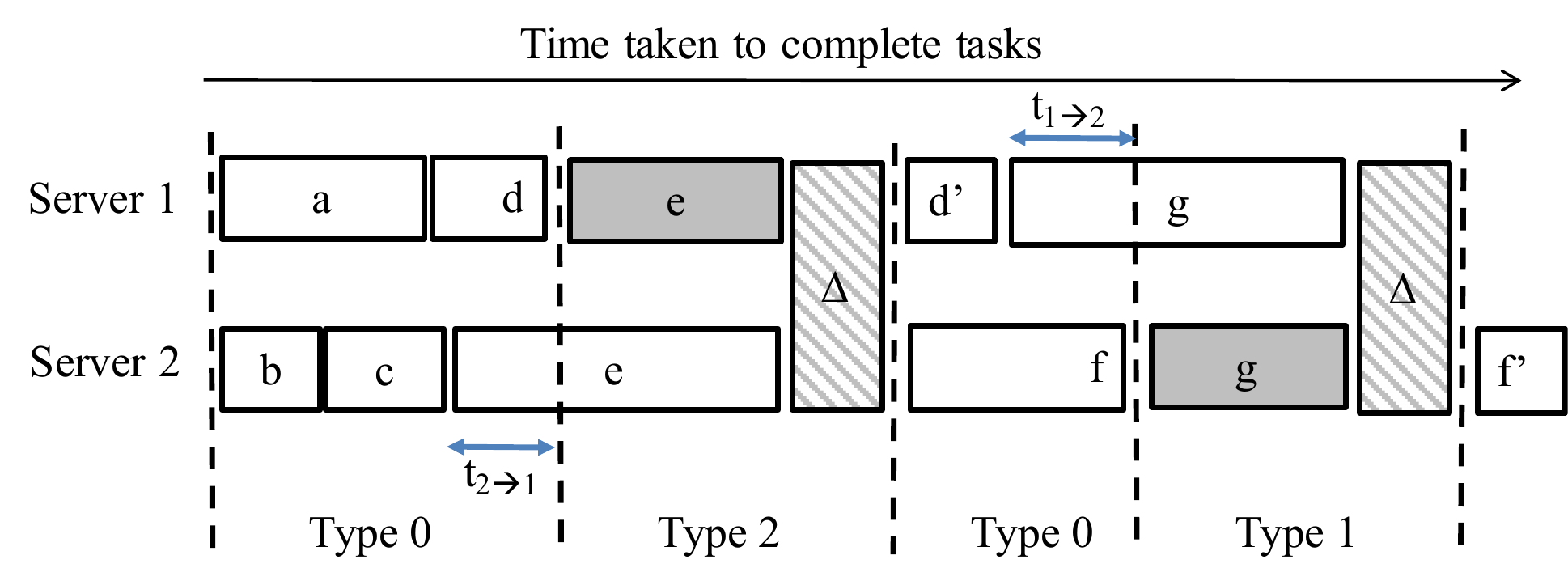}
\caption{Illustration of different types of intervals used to evaluate the throughput in \Cref{thm:two_server_thpt_bnd}. jobs $d$ and $f$ are paused to launch the replicas of $e$ and $g$ respectively, and they are resumed when the replicas are served or canceled.\label{fig:two_server_thpt_bnd}}
\end{figure}

\begin{proof}[Proof of \Cref{thm:two_server_thpt_bnd}]
When $t_{1 \rightarrow 2} = 0$ or $t_{2 \rightarrow 1} = 0$, all jobs are replicated at both servers. Thus by \Cref{lem:thpt_full_rep} we get \eqref{eqn:thpt_bnd_two_server_full_rep}.

Now consider the case where $t_{1 \rightarrow 2} > 0$ and $t_{2 \rightarrow 1} > 0$. Time can be divided into three types of intervals as illustrated in \Cref{fig:two_server_thpt_bnd}. In Type $0$ intervals, no jobs are replicated. In a Type $1$ interval, both servers are serving a job that was originally launched on server $1$. As soon as any one copy finishes, its replica is canceled. Then server $2$ can resume its paused job, and we go back to a Type $0$ interval. Similarly, in a Type $2$ interval, both servers are serving a job that was originally run on server $2$. 

One job departs the system at the end of each Type $1$ or Type $2$ interval. Consider that this departure time is shifted to the end of the Type $0$ preceding this Type $1$ or Type $2$ interval. This shift does not affect the overall throughput. Further, we rearrange the intervals to concatenate all Type $0$ intervals together at the beginning of the time horizon, followed by all Type $1$ and Type $2$ intervals. Now the concatenated Type $0$ interval can be viewed as a system of two servers running jobs according to the no replication policy, with service times $X_1^{tr}(t_{1 \rightarrow 2}) = \min(X_1, t_{1 \rightarrow 2})$ and $X_2^{tr}(t_{2 \rightarrow 1}) = \min(X_2, t_{2 \rightarrow 1})$, which are truncated versions of the original service times. Thus the rate of job completion in the concatenated Type $0$ interval is 
\begin{align}
\Thpt_0 &= \frac{1}{\E{X_1^{tr}(t_{1 \rightarrow 2}) }} + \frac{1}{\E{X_2^{tr}(t_{2 \rightarrow 1}) }}.
\end{align}

Since all job departures are shifted to the end of Type $0$ intervals, the rate of job completion in Type $1$ and Type $2$ intervals is zero, that is, $\Thpt_1 = \Thpt_2 = 0$. The overall throughput can be expressed as
\begin{align}
\Thpt_{p,r} &= \mu_0 \Thpt_0 + \mu_1 \Thpt_1 + \mu_2 \Thpt_2 \label{eqn:thpt_as_sum} \\
&= \mu_0 \Thpt_0 \label{eqn:thpt_in_terms_of_thpt_0},
\end{align}
where $\Thpt_i$ is the rate of job completion in concatenated interval of Type $i$. The weight $\mu_i$ is the fraction of total time spent in a Type $i$ interval. The ratios $\mu_1/\mu_0$ and $\mu_2/\mu_0$ can be expressed in terms of $t_{1 \rightarrow 2}$ and $t_{2 \rightarrow 1}$ as follows.
\begin{align}
\frac{\mu_1}{\mu_0} &= \frac{\Pr(X_1 > t_{1 \rightarrow 2}) (\Delta + \E{\min(X_1^{rs}(t_{1 \rightarrow 2}), X_2)})}{\E{X_1^{tr}(t_{1 \rightarrow 2})}} \label{eqn:mu_1_by_mu_0} \\
\frac{\mu_2}{\mu_0} &= \frac{\Pr(X_2 > t_{2 \rightarrow 1}) (\Delta + \E{\min(X_1, X_2^{rs}(t_{2 \rightarrow 1})})}{\E{X_2^{tr}(t_{2 \rightarrow 1})}} \label{eqn:mu_2_by_mu_0} 
\end{align}

Every job originally run on server $1$ spends $\E{X_1^{tr}(t_{1 \rightarrow 2})}$ expected time in a Type $0$ interval, and $\Pr(X_1 > t_{1 \rightarrow 2}) (\Delta + \E{\min(X_1^{rs}(t_{1 \rightarrow 2}), X_2)})$ expected time in a Type $1$ interval. Thus, the ratio $\mu_1/\mu_0$ is given by \eqref{eqn:mu_1_by_mu_0}. Similarly we get \eqref{eqn:mu_2_by_mu_0}.

Using \eqref{eqn:mu_1_by_mu_0} and \eqref{eqn:mu_2_by_mu_0} along with the fact that $\mu_0 + \mu_1 + \mu_2 = 1$, we can solve for $\mu_i$. Substituting $\mu_0$ in \eqref{eqn:thpt_in_terms_of_thpt_0}, we get the result in \eqref{eqn:thpt_bnd_two_server}. 

\end{proof}

\begin{proof}[Proof of \Cref{lem:single_pt_optimality}]
AdaRep policies replicate a job run on server $1$ (or server $2$) after a fixed elapsed time $t_{1 \rightarrow 2}$ (or respectively $t_{2 \rightarrow 1}$). Instead of fixed $\mathbf{t}$, the replication thresholds could be chosen randomly such that the threshold vector $\mathbf{t}^{(i)}$ for some $i \in [1, 2, \dots I]$ is chosen with probability $Pr(\mathbf{t} = \mathbf{t}^{(i)})$. First let us show that this does not improve the throughput.

We can divide time into $I$ types of intervals, such that in the Type $i$ interval, replicas are launched according to the threshold vector $\mathbf{t}^{(i)}$. We can concatenate all intervals of Type $i$ together. Each type $i$ interval can be further divided into three types sub-intervals as given in the proof of \Cref{thm:two_server_thpt_bnd} to compute the rate of job completion in that interval. The overall throughput can be expressed as a linear combination of rates of job completion in each of these interval types,
\begin{align}
R_{p,r} &= \sum_{i=1}^{I} Pr(\mathbf{t} = \mathbf{t}^{(i)}) R_{p,r} (\mathbf{t}^{(i)}) \label{eqn:thpt_as_time_share} \\
&\leq \sum_{i=1}^{I} Pr(\mathbf{t} = \mathbf{t}^{(i)}) \max_{\mathbf{t}} R_{p,r}(\mathbf{t})\\
&= \max_{\mathbf{t}} R_{p,r}(\mathbf{t})
\end{align}
where $Pr(\mathbf{t} = \mathbf{t}^{(i)})$ is the fraction of time spent in the Type $i$ interval. 
The throughput of the best fixed threshold policy upper bounds each term in \eqref{eqn:thpt_as_time_share}. 

At any time instant the scheduler has the information of the elapsed times of the job that is to the replicated and that of the job that will be paused. AdaRep policies only consider the elapsed time of the job to be replicated. We now show that considering the elapsed time of the job that will be paused does not improve the throughput. For the two server case, there are two candidates for replication, the job at server $1$ with an elapsed time $t_{1 \rightarrow 2}$ and the job at server $2$ with an elapsed time $t_{2 \rightarrow 1}$. If the job at server $1$ (job at server $2$) is replicated, then the elapsed time of the job that will be paused is $t_{2 \rightarrow 1}$ ($t_{1 \rightarrow 2}$), which is already considered in the AdaRep policies replication thresholds. % at servers the current running jobs include  of the jobs that are candidates of replication To prove this we show that the throughput of any scheduling policy is independent of the elapsed time of the paused job. For any scheduling policy, the time horizon can be divided into three types of intervals as shown \Cref{fig:two_server_thpt_bnd}. Consider that the departures at the end of Type $1$ and $2$ are shifted to the end of the preceding Type $0$ intervals. From the throughput analysis in the proof of \Cref{thm:two_server_thpt_bnd} we can see that the rate of job completion in the concatenated Type $0$ interval, and the fraction of time $\mu_0$ only depend on the elapsed times $t_{1 \rightarrow 2}$ and $t_{2 \rightarrow 1}$. 
Thus, considering the elapsed times of the job to be paused does not improve the throughput in the two-server case. Therefore, there is no loss of generality in restricting our attention to the class of AdaRep policies. However, this may not be true for the case of $3$ or more servers.
\end{proof}

\begin{proof}[Proof of \Cref{thm:general_upper_bnd}]
Consider a job that enters the central queue and is served by the system of $K$ homogeneous servers by launching copies at one or more servers. Without loss of generality, suppose that the first copy of a job starts at time $t_1 = 0$. Relative to this time, up to $K-1$ additional replicas start at times $t_2 \leq t_3 \leq  \dots \leq t_K$ respectively. If only $r < K$ copies of the job are launched, then  $t_{r+1}$, \dots $t_K$ are $\infty$. As soon as any one of these replicas is served, the others are canceled.

We first express the computation cost $C(t_2, t_3, \dots, t_K)$, that is, the total time collectively spent by the $K$ servers on this job, in terms of these relative start times of the replicas $t_2 \leq t_3 \leq  \dots \leq t_K$. First, observe that the service time $S$ of the job, that is, the time from when the original copy of the job starts until the earliest replica finishes can be expressed in terms of the task start times as $S = \min(X_1, X_2+ t_2, \dots,  X_K+t_K)$. Since the $k^{th}$ server starts executing a replica of the job at time $t_i$ and the replicas are canceled at time $S$, the $k^{th}$ server spends $(S-t_k)^+$ on the job. Therefore, the total computation time (not including cancellation delay) collectively spent by the $K$ servers on the job is  $\sum_{k=1}^{K} (S-t_k)^+$. A cancellation delay $\Delta$ is incurred at all the servers where the job has started service (including the server that finishes first). But if the job is launched at only one server and it finishes before any replica starts ($S < t_2$) then there is no cancellation delay. Therefore, the expected computation time (including cancellation delay) is given by the expression \eqref{eqn:C_in_terms_of_t_k}.

The relative start times of the replicas $t_2 \leq t_3 \leq  \dots \leq t_K$ depend on the choice of the replication policy (for example MaxRate, AdaRep, upfront replication etc.) as well as service times of previous jobs which determine when the servers become available to serve current job. Thus, the throughput of any replication policy is 
\begin{align}
R_{n,r} = \frac{K}{\mathbb{E}_{t_2, \dots, t_K} \mathbb{E}_X[C(t_2, t_3, \dots, t_K)]}
\end{align}
where the joint distribution of $t_2, \dots, t_K$ depends on the choice of the replication policy. Since expectation is lower-bounded by the minimum, we get the upper bound
\begin{align}
R^*_{n,r} \leq \frac{K}{ \min_{0 \leq t_2 \leq \dots \leq t_{K}} \mathbb{E}_{X}[C(t_2, t_3, \dots, t_K)]}.
\end{align}

\end{proof}

%  Bibliography %

\bibliographystyle{ieeetr}
\bibliography{adarep_refs}

\begin{thebibliography}{10}

\bibitem{joshi2018synergy}
G.~Joshi, ``Synergy via redundancy: Boosting service capacity with adaptive
  replication,'' {\em SIGMETRICS Performance Evaluation Review}, vol.~45,
  pp.~21--28, Mar. 2018.

\bibitem{dean2013tail}
J.~Dean and L.~Barroso, ``{The Tail at Scale},'' {\em {Communications of the
  ACM}}, vol.~56, no.~2, pp.~74--80, 2013.

\bibitem{dean2008mapreduce}
J.~Dean and S.~Ghemawat, ``{MapReduce: simplified data processing on large
  clusters},'' {\em ACM Commun.\ Mag.}, vol.~51, pp.~107--113, Jan. 2008.

\bibitem{zaharia_spark_2010}
M.~Zaharia, M.~Chowdhury, M.~J. Franklin, S.~Shenker, and I.~Stoica, ``Spark:
  cluster computing with working sets,'' in {\em Proceedings of the 2nd USENIX
  conference on Hot topics in cloud computing}, vol.~10, p.~10, 2010.

\bibitem{zaharia_sparrow_2013}
K.~Ousterhout, P.~Wendell, M.~Zaharia, and I.~Stoica, ``Sparrow: Distributed,
  low latency scheduling,'' in {\em Proceedings of the ACM Symposium on
  Operating Systems Principles (SOSP)}, pp.~69--84, 2013.

\bibitem{maxemchuk2}
N.~F. Maxemchuk, ``Dispersity routing,'' {\em Proceedings of the International
  Conference on Communications (ICC)}, pp.~10--13, Jun. 1975.

\bibitem{vulimiri2013low}
A.~Vulimiri, P.~B. Godfrey, R.~Mittal, J.~Sherry, S.~Ratnasamy, and S.~Shenker,
  ``Low latency via redundancy,'' in {\em Proceedings of the {ACM} Conference
  on Emerging Networking Experiments and Technologies (CoNEXT)}, pp.~283--294,
  2013.

\bibitem{koole2008resource}
G.~Koole and R.~Righter, ``Resource allocation in grid computing,'' {\em
  Journal of Scheduling}, vol.~11, pp.~163--173, June 2008.

\bibitem{joshi2014delay}
G.~Joshi, Y.~Liu, and E.~Soljanin, ``{On the Delay-storage Trade-off in Content
  Download from Coded Distributed Storage},'' {\em IEEE Journal on Selected
  Areas on Communications}, May 2014.

\bibitem{shah2016when}
N.~B. Shah, K.~Lee, and K.~Ramchandran, ``When do redundant requests reduce
  latency?,'' {\em {IEEE} Transactions on Communications}, vol.~64, no.~2,
  pp.~715--722, 2016.

\bibitem{sun2015provably}
Y.~Sun, Z.~Zheng, C.~E. Koksal, K.~Kim, and N.~B. Shroff, ``Provably delay
  efficient data retrieving in storage clouds,'' in {\em Proceedings of IEEE
  INFOCOM}, Apr. 2015.

\bibitem{gardner2015reducing}
K.~Gardner, S.~Zbarsky, S.~Doroudi, M.~Harchol-Balter, E.~Hyyti{\"a}, and
  A.~Scheller-Wolf, ``Reducing latency via redundant requests: Exact
  analysis,'' in {\em Proceedings of the {ACM} {SIGMETRICS}}, Jun. 2015.

\bibitem{joshi2015efficient}
G.~Joshi, E.~Soljanin, and G.~Wornell, ``Efficient replication of queued tasks
  for latency reduction in cloud systems,'' in {\em Proceedings of the Allerton
  Conference}, Oct. 2015.

\bibitem{joshi2017efficient}
G.~Joshi, E.~Soljanin, and G.~Wornell, ``Efficient redundancy techniques for
  latency reduction in cloud systems,'' {\em ACM Transactions on Performance
  Evaluation of Computer Systems}, vol.~2, pp.~1--30, Apr. 2017.

\bibitem{joshi2012coding}
G.~Joshi, Y.~Liu, and E.~Soljanin, ``Coding for fast content download,'' in
  {\em Proceedings of the Allerton Conference on Comm., Control and Computing},
  pp.~326--333, Oct. 2012.

\bibitem{kleinrock1975theory}
L.~Kleinrock, {\em Theory, Volume 1, Queueing Systems}.
\newblock New York, NY, USA: Wiley-Interscience, 1975.

\bibitem{mor_book}
M.~Harchol-Balter, {\em Performance Modeling and Design of Computer Systems:
  Queueing Theory in Action}.
\newblock Cambridge University Press, 2013.

\bibitem{powerof2}
M.~Mitzenmacher, {\em {The power of two choices in randomized load balancing}}.
\newblock PhD thesis, University of California Berkeley, CA, 1996.

\bibitem{ananthanarayanan2013effective}
G.~Ananthanarayanan, A.~Ghodsi, S.~Shenker, and I.~Stoica, ``Effective
  straggler mitigation: Attack of the clones,'' in {\em Proceedings of the 10th
  {USENIX} Conference on Networked Systems Design and Implementation},
  pp.~185--198, Apr. 2013.

\bibitem{gardner2016power}
K.~Gardner, S.~Zbarsky, M.~Harchol-Balter, and A.~Scheller-Wolf, ``The power of
  d choices for redundancy,'' in {\em Proceedings of the 2016 ACM SIGMETRICS
  International Conference on Measurement and Modeling of Computer Science},
  (New York, NY, USA), p.~409–410, Association for Computing Machinery, 2016.

\bibitem{gardner2016s&x}
K.~Gardner, M.~Harchol-Balter, and A.~Scheller-Wolf, ``A better model for job
  redundancy: Decoupling server slowdown and job size,'' in {\em Proceedings of
  IEEE MASCOTS}, Sept. 2016.

\bibitem{joshi2015queues}
G.~Joshi, E.~Soljanin, and G.~Wornell, ``Queues with redundancy: Latency-cost
  analysis,'' in {\em Proceedings of the ACM SIGMETRICS Workshop on
  Mathematical Modeling and Analysis}, June 2015.

\bibitem{ayesta2018unifying}
U.~Ayesta, T.~Bodas, and I.~M. Verloop, ``On a unifying product form framework
  for redundancy models,'' {\em Performance Evaluation}, vol.~127-128,
  pp.~93--119, Nov. 2018.

\bibitem{raaijmakers2019delta}
Y.~Raaijmakers, S.~Borst, and O.~Boxma, ``Delta probing policies for
  redundancy,'' {\em SIGMETRICS Performance Evaluation Review}, vol.~46,
  p.~72–73, Jan. 2019.

\bibitem{wang2019efficient}
D.~Wang, G.~Joshi, and G.~W. Wornell, ``Efficient straggler replication in
  large-scale parallel computing,'' {\em ACM Trans. Model. Perform. Eval.
  Comput. Syst.}, vol.~4, Apr. 2019.

\bibitem{aktas2017service}
M.~Aktas, S.~E. Anderson, A.~Johnston, G.~Joshi, S.~Kadhe, G.~L. Matthews,
  C.~Mayer, and E.~Soljanin, ``On the service capacity of accessing erasure
  coded content,'' in {\em Proc. Allerton Conf. Commun., Control and
  Computing}, Oct. 2017.

\bibitem{poloczek2016contrasting}
F.~Poloczek and F.~Ciucu, ``Contrasting effects of replication in parallel
  systems: From overload to underload and back,'' {\em arXiv:1602.07978}, Feb.
  2016.

\bibitem{anton2019stability}
E.~Anton, U.~Ayesta, M.~Jonckheere, and I.~M. Verloop, ``On the stability of
  redundancy models,'' 2019.

\bibitem{raaijmakers2019redundancy}
Y.~Raaijmakers, S.~Borst, and O.~Boxma, ``Redundancy scheduling with scaled
  bernoulli service requirements,'' {\em Queueing Systems}, vol.~93, no.~1,
  pp.~67--82, 2019.

\bibitem{anton2020improving}
E.~Anton, U.~Ayesta, M.~Jonckheere, and I.~Verloop, ``Improving the performance
  of heterogeneous data centers through redundancy,'' 2020.

\bibitem{lee2017speeding}
K.~Lee, M.~Lam, R.~Pedarsani, D.~Papailiopoulos, and K.~Ramchandran, ``Speeding
  up distributed machine learning using codes,'' {\em IEEE Transactions on
  Information Theory}, 2017.

\bibitem{yu2017polynomial}
Q.~Yu, M.~Maddah-Ali, and S.~Avestimehr, ``Polynomial codes: an optimal design
  for high-dimensional coded matrix multiplication,'' in {\em Advances in
  Neural Information Processing Systems}, pp.~4406--4416, 2017.

\bibitem{dutta2016short}
S.~Dutta, V.~Cadambe, and P.~Grover, ``Short-dot: Computing large linear
  transforms distributedly using coded short dot products,'' in {\em Advances
  In Neural Information Processing Systems}, pp.~2100--2108, 2016.

\bibitem{kosaian2019parity}
J.~Kosaian, K.~V. Rashmi, and S.~Venkataraman, ``Parity models: {A} general
  framework for coding-based resilience in {ML} inference,'' {\em CoRR},
  vol.~abs/1905.00863, 2019.

\bibitem{mallick2019rateless}
A.~Mallick and G.~Joshi, ``Rateless codes for distributed computations with
  sparse compressed matrices,'' in {\em IEEE International Symposium on
  Information Theory (ISIT)}, jul 2019.

\bibitem{mallick2020rateless}
A.~Mallick, U.~Sheth, G.~Palanikumar, M.~Chaudhari, and G.~Joshi, ``{Rateless
  Codes for Near-Perfect Load Balancing in Distributed Matrix-Vector
  Multiplication},'' in {\em {ACM Sigmetrics 2020}}, May 2020.

\bibitem{varki_merc_chen}
E.~Varki, A.~Merchant, and H.~Chen, ``The {M/M/1} fork-join queue with variable
  sub-tasks,'' {\em {unpublished, available online}}, 2008.

\bibitem{nelson_tantawi}
R.~Nelson and A.~Tantawi, ``Approximate analysis of fork/join synchronization
  in parallel queues,'' {\em IEEE Transactions on Computers}, vol.~37,
  pp.~739--743, Jun. 1988.

\bibitem{rizk_poloczek_ciucu_2015}
A.~Rizk, F.~Poloczek, and F.~Ciucu, ``Computable bounds in fork-join queueing
  systems,'' in {\em Proceedings of the 2015 ACM SIGMETRICS International
  Conference on Measurement and Modeling of Computer Systems}, SIGMETRICS
  ’15, (New York, NY, USA), p.~335–346, Association for Computing
  Machinery, 2015.

\bibitem{xiang2016joint}
Y.~{Xiang}, T.~{Lan}, V.~{Aggarwal}, and Y.~R. {Chen}, ``Joint latency and cost
  optimization for erasure-coded data center storage,'' {\em IEEE/ACM
  Transactions on Networking}, vol.~24, no.~4, pp.~2443--2457, 2016.

\bibitem{parag2017latency}
P.~Parag, A.~Bura, and J.-F. Chamberland, ``Latency analysis for distributed
  storage,'' in {\em IEEE International Conference on Computer Communications
  (INFOCOM)}, May 2017.

\bibitem{badita2019latency}
A.~{Badita}, P.~{Parag}, and J.~{Chamberland}, ``Latency analysis for
  distributed coded storage systems,'' {\em IEEE Transactions on Information
  Theory}, vol.~65, pp.~4683--4698, Aug 2019.

\bibitem{li_mean_field_2016}
B.~Li, A.~Ramamoorthy, and R.~Srikant, ``Mean-field-analysis of coding versus
  replication in cloud storage systems,'' in {\em IEEE International Conference
  on Computer Communications (INFOCOM)}, pp.~1--9, April 2016.

\bibitem{aktas2020service}
M.~Aktas, G.~Joshi, S.~Kadhe, F.~Kazemi, and E.~Soljanin, ``Service rate
  region: A new aspect of coded distributed system design,'' 2020.

\end{thebibliography}

\vfill{}
\end{document}